\documentclass[%
 reprint,
 amsmath,amssymb,
 aps,
prl,
]{revtex4-1}

\usepackage{xcolor}
\usepackage{graphicx}
\usepackage{dcolumn}
\usepackage{bm}
\usepackage{hyperref}

\hypersetup{
  colorlinks=true
}

\begin{document}

\title{Prediction and manipulation of hydrodynamic rogue waves via nonlinear spectral engineering}

\author{Alexey Tikan}
\altaffiliation{Univ. Lille, CNRS, UMR 8523 - PhLAM -  Physique des Lasers Atomes et Mol\'ecules, F-59000 Lille, France}
\altaffiliation{Present address: Institute of Physics, Swiss Federal Institute of Technology Lausanne (EPFL), CH-1015 Lausanne, Switzerland}
\email{alexey.tikan@epfl.ch}

\author{Felicien Bonnefoy}
\altaffiliation{\'Ecole Centrale de Nantes, LHEEA, UMR 6598 CNRS, F-44 321 Nantes, France}

\author{Giacomo Roberti}
\altaffiliation{Department of Mathematics, Physics and Electrical Engineering, Northumbria University, Newcastle upon Tyne, NE1 8ST, United Kingdom}

\author{Gennady El}
\altaffiliation{Department of Mathematics, Physics and Electrical Engineering, Northumbria University, Newcastle upon Tyne, NE1 8ST, United Kingdom}

\author{Alexander Tovbis}
\altaffiliation{Department of Mathematics, University of Central Florida, Orlando, Florida, 32816, USA}

\author{Guillaume Ducrozet}
\altaffiliation{\'Ecole Centrale de Nantes, LHEEA, UMR 6598 CNRS, F-44 321 Nantes, France}

\author{Annette Cazaubiel}
\altaffiliation{Universit\'e de Paris, MSC, UMR 7057 CNRS, F-75 013 Paris, France}

\author{Gaurav Prabhudesai}
\altaffiliation{Laboratoire de Physique de l’\'Ecole normale superieure, ENS, Universite PSL, CNRS, Sorbonne Universit\'e, Universit\'e Paris-Diderot, 75005 Paris, France}

\author{Guillaume Michel}
\altaffiliation{Institut Jean Le Rond d’Alembert, Sorbonne Universit\'e, CNRS, UMR 7190, F-75 005 Paris, France}

\author{Francois Copie}
\altaffiliation{Univ. Lille, CNRS, UMR 8523 - PhLAM -  Physique des Lasers Atomes et Mol\'ecules, F-59000 Lille, France}

\author{Eric Falcon}
\altaffiliation{Universit\'e de Paris, MSC, UMR 7057 CNRS, F-75 013 Paris, France}

\author{Stephane Randoux}
\altaffiliation{Univ. Lille, CNRS, UMR 8523 - PhLAM -  Physique des Lasers Atomes et Mol\'ecules, F-59000 Lille, France}

\author{Pierre Suret}
\altaffiliation{Univ. Lille, CNRS, UMR 8523 - PhLAM -  Physique des Lasers Atomes et Mol\'ecules, F-59000 Lille, France}

\date{\today}

\begin{abstract}
Peregrine soliton (PS) is widely regarded as a prototype nonlinear structure capturing properties of rogue waves that emerge in the nonlinear propagation of unidirectional wave trains. As an exact breather solution of the one-dimensional focusing nonlinear Schr\"odinger equation with nonzero boundary conditions, the PS can be viewed as a soliton on finite background, {\it i.e.} a nonlinear superposition of a soliton and a monochromatic wave. A recent mathematical work showed that both nonzero boundary conditions and solitonic content are not pre-requisites for the PS occurrence. Instead, it has been demonstrated that PS can emerge {\it locally}, as an asymptotic structure arising from the propagation of an arbitrary large decaying pulse, independently of its solitonic content. This mathematical discovery has changed the widely accepted paradigm of the solitonic nature of rogue waves by enabling the PS to emerge from a partially radiative or even completely solitonless initial data. 
In this work, we realize the mathematically predicted universal mechanism of the local PS emergence in a water tank experiment with a particular aim to control the point of the PS occurrence in space-time by imposing an appropriately chosen initial chirp.
By employing the inverse scattering transform for the synthesis of the initial data, we are able to engineer a localized wave packet with a prescribed solitonic and radiative content. This enabled us to control the position of the emergence of the rogue wave by adjusting the inverse scattering spectrum. The proposed method of the nonlinear spectral engineering is found to be robust to higher-order nonlinear effects inevitable in realistic wave propagation conditions. 
\end{abstract}

\pacs{Valid PACS appear here}

\maketitle


\section{Introduction}

The prediction of extreme events in various nonlinear media has been the subject of a very active research for several decades~\cite{Annenkov2001predictability,kharif2008Rogue,Onorato2013Rogue,Cousins2014Quantification,Birkholz2015Predictability,Cousins2019Predicting}. While the original motivation was related to the modeling of the emergence of giant water waves in the ocean, also called rogue waves (RWs)~\cite{kharif2008Rogue}, the subsequent research revealed that RWs are a fundamental and ubiquitous physical phenomenon occurring, apart from classical fluids, in
optical media and superfluids~\cite{Onorato2013Rogue,dudley2019rogue}.

The RW formation as a physical phenomenon has two inherent aspects: dynamical and statistical. 
As a dynamical object RWs are identified according to physical mechanisms responsible for the amplitude growth and spatiotemporal localization. As a statistical object, RWs are characterized by the deviation of the probability distribution of the random wave field from the one implied by the Gaussian statistics ~\cite{Onorato2004Observation,Wabnitz2016Roadmap,Randoux2016Nonlinear,Koussaifi2017Spontaneous}. Among different contexts of the RW's emergence, nonlinear dynamics of unidirectional waves on the surface of the deep water has been widely studied due to \emph{integrability}~\cite{Zakharov:1972Exact} of the mathematical model describing the weakly nonlinear deep-water waves at leading order. Indeed, as has been shown by V. Zakharov~\cite{Zakharov1968Stability}, the evolution of the narrow-band wave packets in the limit of infinite water depth is governed by the one-dimensional focusing nonlinear Schr\"odinger equation (1-D fNLSe), which can be integrated by means of the inverse scattering transform (IST) also known as nonlinear Fourier transform~\cite{novikov1984theory}.

The research area that considers nonlinear evolution of random waves in integrable models has been dubbed \emph{integrable turbulence} in~\cite{Zakharov2009turbulence}. There are two types of initial conditions usually considered in this framework: (i) a plane wave (condensate) perturbed by a small-amplitude noise and (ii) a partially coherent wave with large-scale finite-amplitude variations characterized by long-range incoherence. The latter can be viewed as an infinite sequence of large-scale pulses randomly distributed along the line. The variation of the statistical properties of the IST spectrum occurring in the transition between these two contrasting types of initial conditions has been examined in~\cite{Soto-Crespo2016Integrable,Akhmediev2016Breather}. The random nonlinear wave field (the integrable turbulence) generated in both cases exhibits deviations from the Gaussian statistics~\cite{Agafontsev2015Integrable,kraych2019statistical,copie2019physics,Walczak2015Optical,Suret2016Single,Suret:2017book} but those deviations are manifested at different stages of the evolution. While in the development of the noise-induced modulational instability the non-Gaussian features are observed at the initial stage of the evolution and are of transient character, the nonlinear evolution of partially coherent initial conditions leads to the long-time statistically stationary non-Gaussian state ~\cite{Onorato2004Observation,Walczak2015Optical}. The non-Gaussianity of the asymptotic state in the evolution of partially coherent wave (the so-called heavy tailed distribution) is associated with the presence of RWs. Exact breather solutions of the 1-D fNLSe, also called solitons on finite background~\cite{kuznetsov1977solitons,Ma1979Perturbed,peregrine1983water,akhmediev1986modulation}, are often considered as the main candidates for the role of RWs~\cite{Shrira2010What,Dysthe1999Note,Dudley2014Instabilities}. The PS~\cite{peregrine1983water,Kibler2010Peregrine,Chabchoub2011Rogue} is a particular member of this family of solutions that exhibits spatiotemporal localization, thereby reflecting the main qualitative features of RWs~\cite{Akhmediev2009Waves}. 

The PS solution plays a special role in the context of the dynamics of partially coherent waves. In the strongly nonlinear  regime of propagation, the individual large-scale pulses forming  a partially coherent wave undergo self-focusing resulting in a gradient catastrophe, a  phenomenon of the occurrence of infinite derivatives in the wave's profile. Importantly, the initial, self-focusing stage of the evolution of a partially coherent wave is dominated by  nonlinearity and is approximately dispersionless. For the 1-D fNLSe, it was rigorously proved by Bertola and Tovbis (BT) ~\cite{Bertola2013Universality} that the gradient catastrophe is \emph{universally} regularized by dispersive effects via the local emergence of a coherent structure, which is asymptotically described by the PS solution in the semiclassical, small-dispersion limit. The universality of the PS generation is understood in the sense that this regularization mechanism persists regardless of the particular shape, chirp or \emph{solitonic content} of the initial condition. In particular, the gradient catastrophe regularization via the PS formation for smooth, rapidly decaying purely solitonic initial conditions have been examined experimentally using the optical fiber platform~\cite{Tikan2017Universality,Tikan2021Local}, which revealed the robustness of the mechanism even when typical nonlinear and dispersive lengths are of the same order of magnitude. 

Recently, it has been shown that the local emergence of PSs leaves a distinct trace in the evolution of statistical properties of the partially coherent waves. Numerical simulations performed in~\cite{Tikan2020Effect} have demonstrated that the width and the position of the computed probability density function of local PS emergence positions coincides with the parameters of the \emph{most probable distance interval for the RW observation}, which has been subsequently confirmed in water tank experiments~\cite{Michel2020Emergence}. 
Importantly, the above optics and hydrodynamics experimental results were obtained in the carefully chosen propagation regimes well approximated by the 1-D fNLSe

While having been confirmed experimentally for approximately integrable propagation regimes, the applicability of the universal PS regularization scenario to more physically realistic conditions remains an open question. Indeed, the presence of dissipation and the influence of the higher-order nonlinear terms can significantly modify the integrable 1-D fNLSe dynamics. Having an experimental confirmation of the robustness of the core universal PS resolution dynamics in a physically realistic context would be extremely valuable as it would open a way to a practical, quantitative prediction and manipulation of the {\it spontaneous emergence} of RWs via the methods of nonlinear spectral (IST) theory underlying the BT results. Of course, special care should be taken in the interpretation of the IST data in the context of perturbed integrable dynamics. 

In this paper, using the BT theory as a starting point, we experimentally demonstrate the universality of the spontaneous emergence of the PS-like coherent structures in the evolution of weakly nonlinear wave packets on deep water. As the simplest mathematical model that has an integrable core and takes into account the higher-order nonlinear effects in the propagation of weakly nonlinear wave packets on deep water, we use a version of the generalized fNLSe proposed in~\cite{goullet2011numerical} which is usually referred to as the Dysthe equation (see Methods section). 

The experiments are performed in a 120~$\mathrm{m}$ long water tank with a water depth $h=3 \, \mathrm{m}$. We provide an experimental confirmation of the robustness the BT scenario of the PS emergence in the context of `non-integrable' deep-water wave propagation. It is done by showing that the generation of a RW having the structure similar to PS occurs locally, in the vicinity of the point of the gradient catastrophe and independently of the IST solitonic content of the initial pulse. In particular, we demonstrate that, somewhat paradoxically (but in full agreement with the BT results) the PS-type RW generation is observed even in a completely solitonless case. 

Further, by changing the chirp of the initial wave packet and hence, varying its solitonic content as defined by the IST, we managed to manipulate the point of the PS maximum compression in a controllable way. Our experimental results are shown to be in excellent agreement with the numerical simulations of the Dysthe equation. 
For the analysis of the RW emergence in the wavepacket propagation modeled by the Dysthe equation we employ the recently developed IST based method ~\cite{Bonnefoy2020modulational} of the analysis of nonlinear wave dynamics applicable to the models which are not integrable but have an integrable 'core'~\cite{ryczkowski2018real,Chekhovskoy2019Nonlinear,sugavanam2019analysis,turitsyn2020nonlinear}.

\section{Results}

\subsection{Semi-classical limit of 1-D fNLSe. Universal formation of the Peregrine soliton}

We write the 1-D fNLSe in the following form:
\begin{equation}
i \epsilon \frac{\partial \psi}{\partial \xi} +\frac{ \epsilon^{2}}{2} \frac{\partial^{2} \psi}{\partial \tau^{2}} + |\psi|^{2}\psi = 0, 
\label{eq:NLSeps}
\end{equation}
where $\psi$ is the normalized complex envelope of the water waves, $\xi$ and $\tau$ are normalized space coordinate and normalized time, $\epsilon = \sqrt{L_{NL}/L_D}$, with $L_{NL}$ and $L_D$ typical nonlinear and linear lengths in the system~\cite{Koussaifi2017Spontaneous}.

Despite the fact that 1-D fNLSe is a fully integrable equation~\cite{Zakharov:1972Exact}, explicit analytic results are available only in some particular cases. One of the examples when an effective analytical description is possible is the so-called semi-classical limit of 1-D fNLSe. This approach is applied to the description of nonlinear dispersive waves in the case when $\epsilon \ll 1$ in Eq.~\eqref{eq:NLSeps}.

We consider Equation~\eqref{eq:NLSeps} using the Madelung transformation~\cite{Madelung1927Quantum,El2016Dispersive}:
\begin{equation}
\psi\left ( \xi , \tau\right ) = \sqrt{\rho \left ( \xi , \tau\right )} e^{i \phi \left ( \xi, \tau\right ) /\epsilon}, \; u\left ( \xi , \tau\right ) = \phi_{\tau}\left ( \xi, \tau\right ), 
\end{equation}
where $ \sqrt{\rho}$ is the wave amplitude and $u$--the instantaneous frequency.
As a result the 1-D fNLSe assumes the form of a system
\begin{eqnarray}
\label{eq:WNLS1}
\rho_{\xi} +(\rho u)_{\tau}=0 \\
\label{eq:WNLS2}
u_{\xi} + uu_{\tau} - \rho_{\tau}  + \frac{\epsilon^{2}}{4} \left[ \frac{\rho_{\tau}^{2}}{2\rho^{2}} - \frac{\rho_{\tau \tau}}{\rho} \right]_{\tau}  = 0.
\end{eqnarray}
Equations \eqref{eq:WNLS1}, \eqref{eq:WNLS2} are analogous to Euler's dispersive hydrodynamics for the fluid with density $\rho$ and velocity $u$ but characterized by negative pressure $p = -\rho^{2}/2$ due to the focusing effects.

The last term in Eq.~\eqref{eq:WNLS2} is proportional to $\epsilon^2$, therefore, assuming finite derivatives in the initial data, it can be neglected at the early stage of the propagation. The description of the evolution of smooth and decaying initial profiles by Eqs.~\eqref{eq:WNLS1} and \eqref{eq:WNLS2} with neglected dispersive terms, i.e. with $\epsilon=0$, is valid until the gradients of $\rho$ or $u$ become infinitely large at some point $(\tau_c, \xi_c)$, termed the gradient catastrophe point.  The dispersionless evolution problem is ill-posed for $\xi>\xi_c$, and the full dispersive system has to be considered in this region. To describe the solution in the vicinity of the gradient catastrophe point $(\tau_c, \xi_c)$ BT~\cite{Bertola2013Universality} employed the inverse scattering transform in the semi-classical ($\epsilon \ll 1$) approximation. It has been found that the gradient catastrophe is dispersively regularized by the \textit{universal} appearance of a large amplitude $\epsilon$-scaled spikes that are asymptotically described by the PS solution. Here the term `universal' is used to stress that the BT scenario does not depend on the exact shape, chirp or solitonic content of the smooth (more precisely, analytic) initial condition. It is remarkable that locations of these spikes are determined by the poles of the special tritronqu\'ee solution of the Painlev\'e I equation, whose role in the gradient catastrophe was first recognized by B. Dubrovin et al.~\cite{Dubrovin2009universality}.

\begin{figure}
\center{\includegraphics[width=0.99\linewidth]{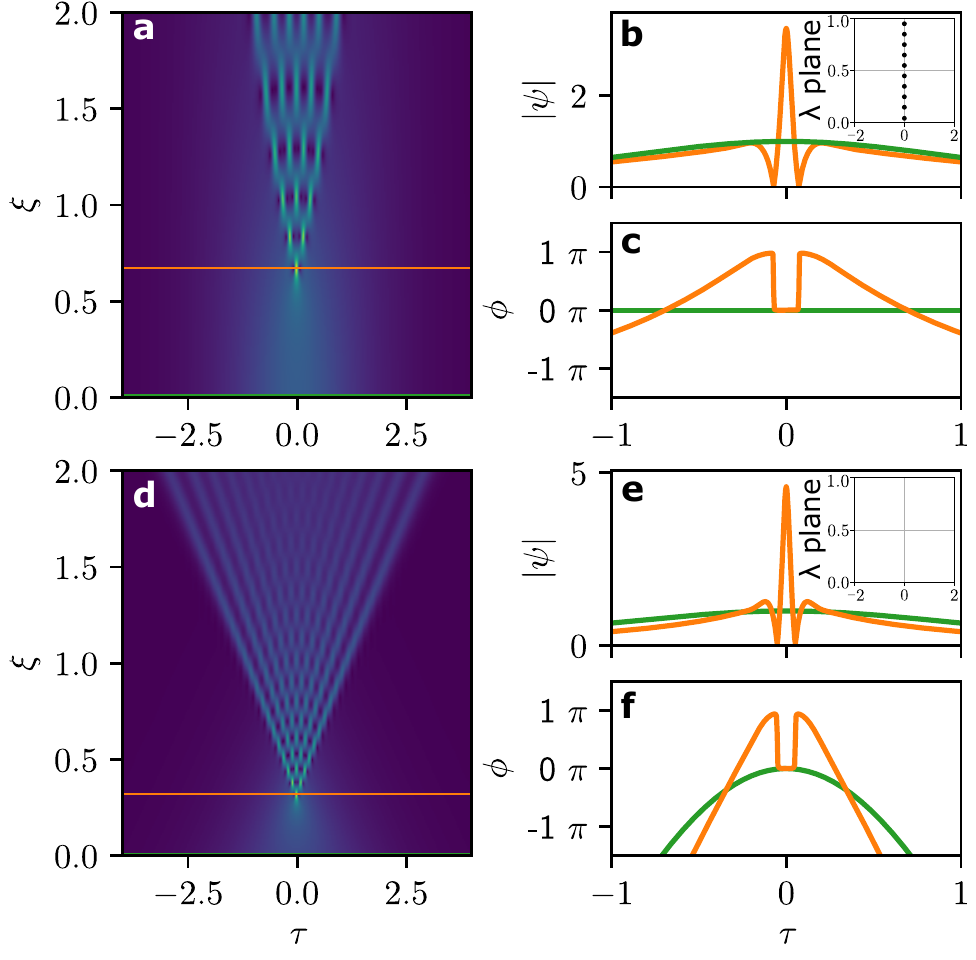}}
\caption{\textbf{Universal regularization of the gradient catastrophe by local emergence of the Peregrine soliton.} Parameter $\epsilon$ in the simulation is equal to 1/10. \textbf{(a-c)} Spatiotemporal diagram, the amplitude and phase cross-section at the maximum compression point (orange). Exact $10$-soliton solution is taken as an initial condition (green). \textbf{(d-f)} The counterpart plots for a solitonless solution $\mathrm{sech}(\tau)\exp[-i\mu \log(\cosh(\tau))/\epsilon]$, where $\mu = 2$. Plots in the right corner of (b) and (e) show the corresponding discrete IST spectra ($\lambda$ plane). Vertical axis shows the imaginary part while the horizontal one - the real part of the discrete IST spectrum.}
\label{fig:1}
\end{figure}

\begin{figure}
\center{\includegraphics[width=0.9\linewidth]{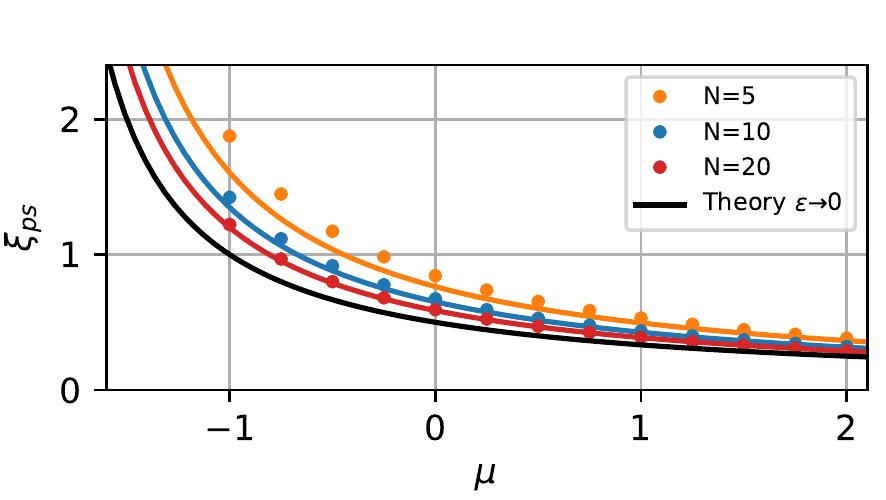}}
\caption{\textbf{The PS emergence distance as a function of the chirp parameter $\mu$. Comparison of the numerical simulations with the theoretical predictions.} Numerical simulations of N = 5, 10 and 20 solitons solutions of the 1-D fNLSe are depicted by orange, blue and red dots respectively. Theoretical prediction according to Eq.~(\ref{eq:PScompressionPoint}) are plotted with the solid line of the corresponding color. The limiting case $N \to \infty$ is plotted with the black line.}
\label{fig:2}
\end{figure}

More explicitly, the theory developed in~\cite{Bertola2013Universality} provides the following asymptotic description of the structure that emerges at the point of maximum compression $\xi=\xi_m$:
\begin{equation}
\label{eq:PScompressionPoint}
\xi_{m} = \xi_{c} + C\epsilon^{4/5},
\end{equation} 
where $\xi_{c}$ is the point of gradient catastrophe,  and $C = \left(\frac{5|C_1|}{4}\right)^{1/5}(2b_0)^{-3/2}|v_p|(1+\mathcal{O}(\epsilon^{4/5}))$ with $C_1$ - an initial data-dependent coefficient, $b_0 = \sqrt{\rho(0,\xi_c)}$ - the wave amplitude at the point of gradient catastrophe, and $|v_p|\approx2.38$ is a universal constant. 
 
The amplitude profile of the spike that emerges at $\xi=\xi_m$ is described by the asymptotic formula:
\begin{equation}
\label{eq:peregrine}
|\psi(\tau, \xi_m)| = a_0 \left(1-\frac{4}{1+4 a_0^2 (\tau/\epsilon)^2} \right)[1+\mathcal{O}(\epsilon^{1/5})],
\end{equation}
which coincides at leading order with the formula of the PS amplitude profile. Here the background amplitude $a_0 = b_0 + \mathcal{O}(\epsilon^{1/5})$, i.e. is determined at leading order by the wave amplitude at the gradient catastrophe point.
  The maximum value of $|\psi|$ in the local PS is then $3a_0$ and is determined up to $\mathcal{O}(\epsilon^{1/5})$. The expression for the phase of the asymptotic solution at the maximum compression point also coincides at leading order with the PS's phase but we do not present it here. We stress that the described approximate PS solution is valid {\it locally}, in the $\epsilon$-vicinity of the point $(0, \xi_{m})$

The local emergence of the PS in the evolution of a decaying $N$-soliton pulse has been experimentally observed in~\cite{Tikan2017Universality,Tikan2021Local}. Importantly, the fiber optics experiments in~\cite{Tikan2017Universality} demonstrated that the BT scenario is very robust and can be observed for the values of $\epsilon$ significantly exceeding those implied by the formal asymptotic validity of the semi-classical limit of 1-D fNLSe. Specifically, the regularization of the gradient catastrophe by the emergence of a coherent structure that can be \textit{locally} fitted by the PS was observed in ~\cite{Tikan2017Universality} for $\epsilon \approx 0.45$ using high-power pulses with a Gaussian profile and a constant phase.

The universality of the described gradient catastrophe regularization mechanism can be illustrated as follows. 
Consider the 1-D fNLSe Eq.~\eqref{eq:NLSeps} with the initial condition
\begin{equation}
\label{eq:sltlss_ic}
 \psi(0,\tau)= \hbox{sech}(\tau) e^{i \phi/\epsilon}, \ \ \phi = -\mu \log(\cosh(\tau)) 
\end{equation}
for two different values of the chirp parameter $\mu$. In the particular case of initial data Eq.~\eqref{eq:sltlss_ic}, coefficients found in Eq.~\eqref{eq:PScompressionPoint} take the following form: $\xi_{c}=1/(2 + \mu$), $C_1=\frac{32\sqrt{2i}}{15(2+\mu)^{9/4}}$, $b_0=\sqrt{\mu+2}$.
For $\mu=0$ the profile Eq.~\eqref{eq:sltlss_ic} is the exact $N$-soliton solution of the 1-D fNLSe (Eq.~\eqref{eq:NLSeps}) with
$\epsilon=1/N$.
On the other hand, if $\mu \geq 2$ the profile Eq.~\eqref{eq:sltlss_ic} is \emph{completely solitonless}~\cite{tovbis_eigenvalue_2000}.

The spatiotemporal diagrams and the profiles of the amplitude $|\psi|$ and phase $\phi$ at the maximum compression point for the evolution of the above two profiles are shown in Fig.~\ref{fig:1}. In both cases the simulations were performed with $\epsilon=1/10$ and the value of $\mu$ in the second set of simulations was taken to be equal $2$ ensuring the absence of the discrete spectrum. In order to verify the solitonic content of the initial data, the IST spectrum of the initial pulse was evaluated numerically using the Fourier collocation method~\cite{Yang:2010} (see the Methods section). The discrete part of the IST spectrum of the $N$-soliton profile is represented by $N$ discrete eigenvalues (and their conjugates) in the complex $\lambda$-plane, located equidistantly along the imaginary axis (Fig.~\ref{fig:1}(b) inset). The IST spectrum of the solitonless initial condition (Eq.~\eqref{eq:sltlss_ic} with $\mu =2$) contains no discrete part, see Fig.~\ref{fig:1}(e) inset. We note that the IST spectra of Gaussian pulses used in the optical experiments of~\cite{Tikan2017Universality} contained both discrete and continuous spectrum parts.

One can see that, although the solitonic content of the two initial conditions is completely different, in both cases the pulse experiences the gradient catastrophe and the coherent structure that emerges at the maximum compression point has the signature amplitude and phase profiles of the PS. As predicted by Eq.~\eqref{eq:PScompressionPoint}, the maximum compression point is shifted further in the presence of the solitonic content. After the first spike we observe in Fig.~\ref{fig:1}(a) the generation of a growing chain of large-amplitude breathers. The qualitative evolution of the solitonless pulse close to the PS regularization point is similar but the long time behaviour is very different, displaying in the $\xi$,$\tau$-plane an expanding cone filled with small-amplitude dispersive waves, see Fig.~\ref{fig:1}(d).

The predicted accuracy of the PS emergence position by Eq.~\eqref{eq:PScompressionPoint} is examined by numerically simulating Eq.~\eqref{eq:NLSeps} for a series of initial conditions in the form of Eq.~\eqref{eq:sltlss_ic} with different values of the chirp parameter $\mu\in[-1,2]$. The results are displayed in Fig.~\ref{fig:2}, where solid lines show the estimates of the PS emergence distance according to Eq.~\eqref{eq:PScompressionPoint} for 5, 10 and 20 solitons at $\mu=0$ (orange, blue and red colors, respectively) and the asymptotic case of infinite number of solitons is shown by the black curve. The points depict the position of PS emergence found in numerical simulations of Eq.~\eqref{eq:NLSeps} (color correspondence is preserved). Fig.~\ref{fig:2} demonstrates the rapid convergence of the theoretical estimates towards the results of 1-D fNLSe simulations with increasing number of solitons in the initial condition and, importantly, their ability to provide a sufficiently accurate prediction even for low soliton numbers, i.e. way beyond the semiclassical zero-dispersion limit considered in the derivation of Eq.~\eqref{eq:PScompressionPoint}.

The above examples clearly show that, despite the widely accepted paradigm of the `solitonic' nature of the PS, the presence of the discrete IST spectrum in the initial data is not a pre-requisite for the PS emergence.
One can also conclude that the emergence of the local PS as a result of $N$-soliton self-compression observed in \cite{Tikan2017Universality} is just a particular case of the general regularization mechanism described by the BT semiclassical theory.

\subsection{Experimental results and comparison with numerical simulations}

\begin{figure*}
\center{\includegraphics[width=0.6\linewidth]{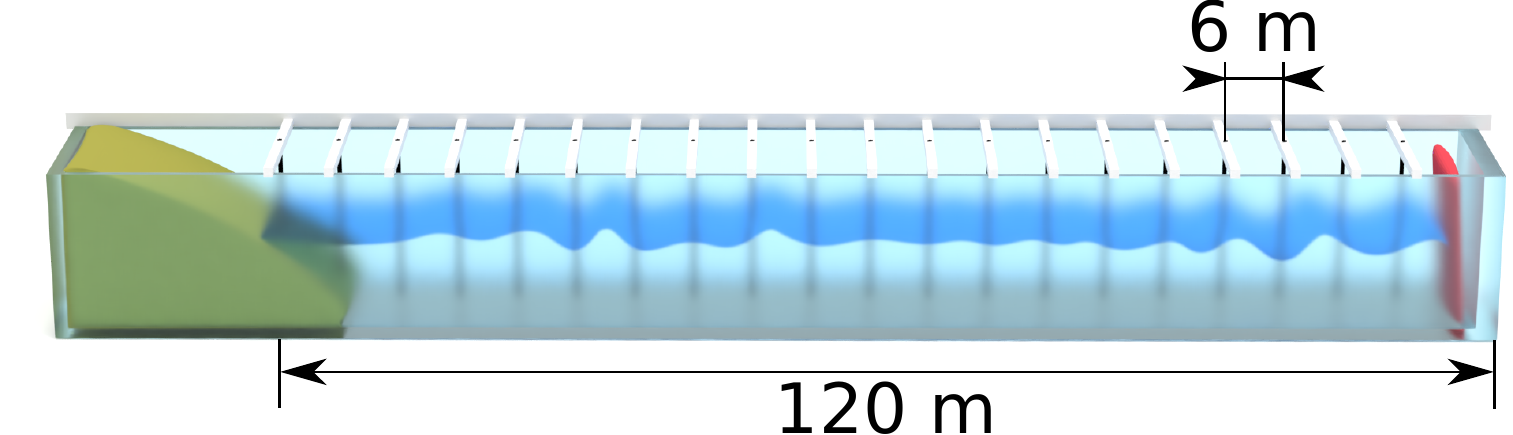}}
\caption{\textbf{Schematic representation of the water tank.} Experimental investigations are provided in the water tank of the Hydrodynamics, Energetics, and Atmospheric Environment Lab (LHEEA) at Ecole Centrale de Nantes (France). Wave elevation is measured by a set of equidistantly-spaced probes over 120~m of the  water tank length (the total length is 148~m), at every 6~m. It is equipped with a parabolic shaped absorbing beach ($\approx$8~m long). With the addition of pool lanes arranged in a W pattern in front of the beach, the measured amplitude reflection coefficient is as low as 1\%.}
\label{fig:3}
\end{figure*}

\begin{figure*}
\center{\includegraphics[width=0.7\linewidth]{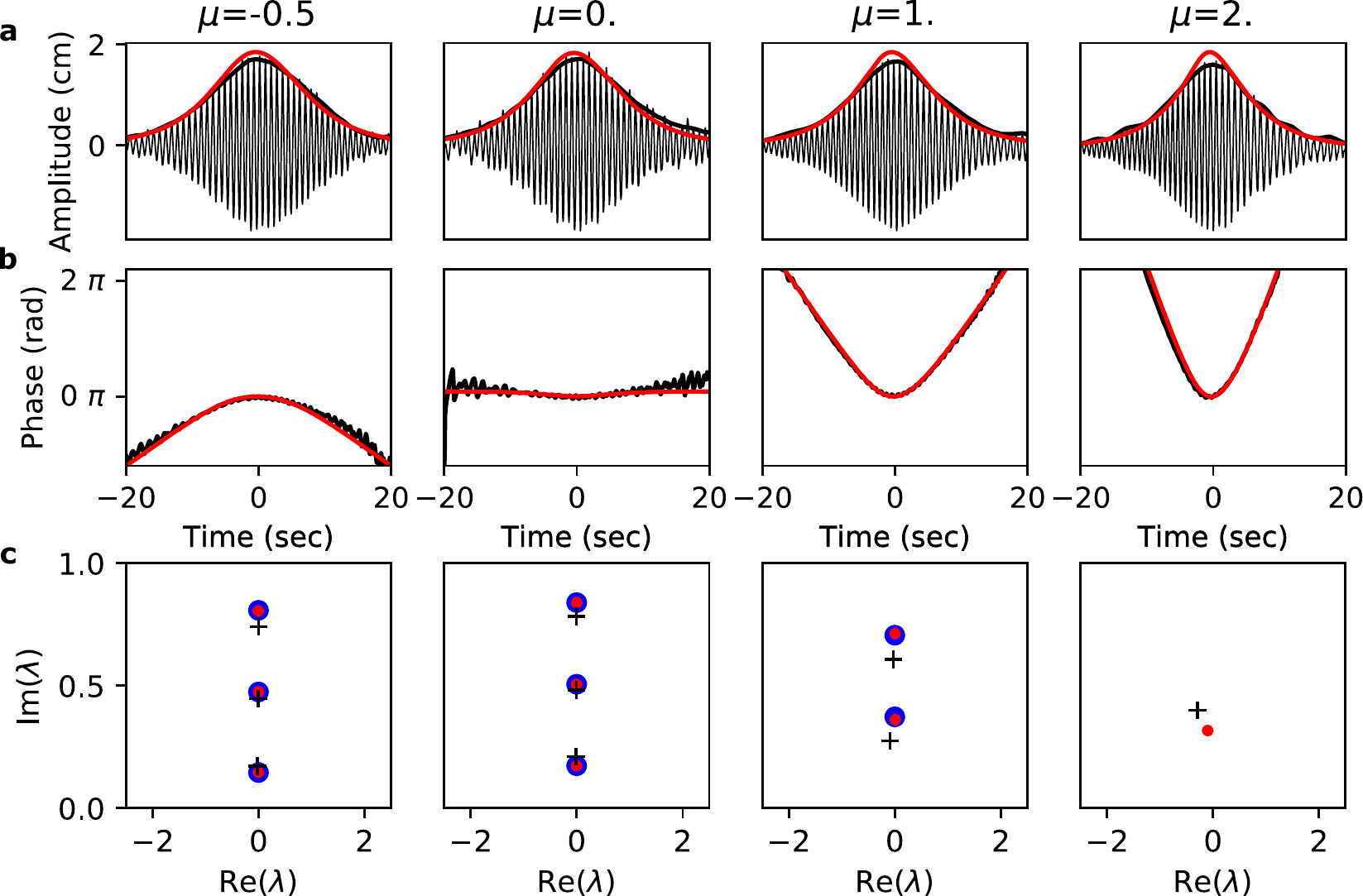}}
\caption{\textbf{Initial conditions for experiments and simulations.} Four columns correspond to the values of $\mu$ = -0.5, 0, 1 and 2, respectively. \textbf{(a)} Comparison of the wave packets measured at the first probe located at 6 m from the wave maker (black line), corresponding envelope (bold black line), and simulations of Dysthe equation (red line) starting from the exact analytical expression. \textbf{(b)} Corresponding phase. \textbf{(c)} Discrete IST spectra of signals measured at the first probe (black crosses), analytical solution (blue dots) and result of simulations of Dysthe equation at 6 m of propagation (red dots).}
\label{fig:cmp_cut_ic}
\end{figure*}

\subsubsection{Experimental parameters and conditions}
We investigate the influence of the soliton content on the emergence of RW in the 120~$\mathrm{m}$ long water tank with a water depth $h=3 \, \mathrm{m}$. To measure the surface wave elevation along the tank, 20 resistive probes have been installed equidistantly with the separation of 6~$\mathrm{m}$. A schematic representation of the water tank is shown is Fig.~\ref{fig:3}. Detailed description of the experimental platform as well as the mode of operation can be found in~\cite{Bonnefoy2020modulational,Michel2020Emergence}. All the experiments are performed in the deep-water regime with a typical value of $kh = $15.8, where $k = 2 \pi / \lambda$ is the wavenumber. The central frequency of the wave packet is set to 1.15 Hz. Additional information can be found in the Methods section.

\begin{figure*}
\center{\includegraphics[width=0.8\linewidth]{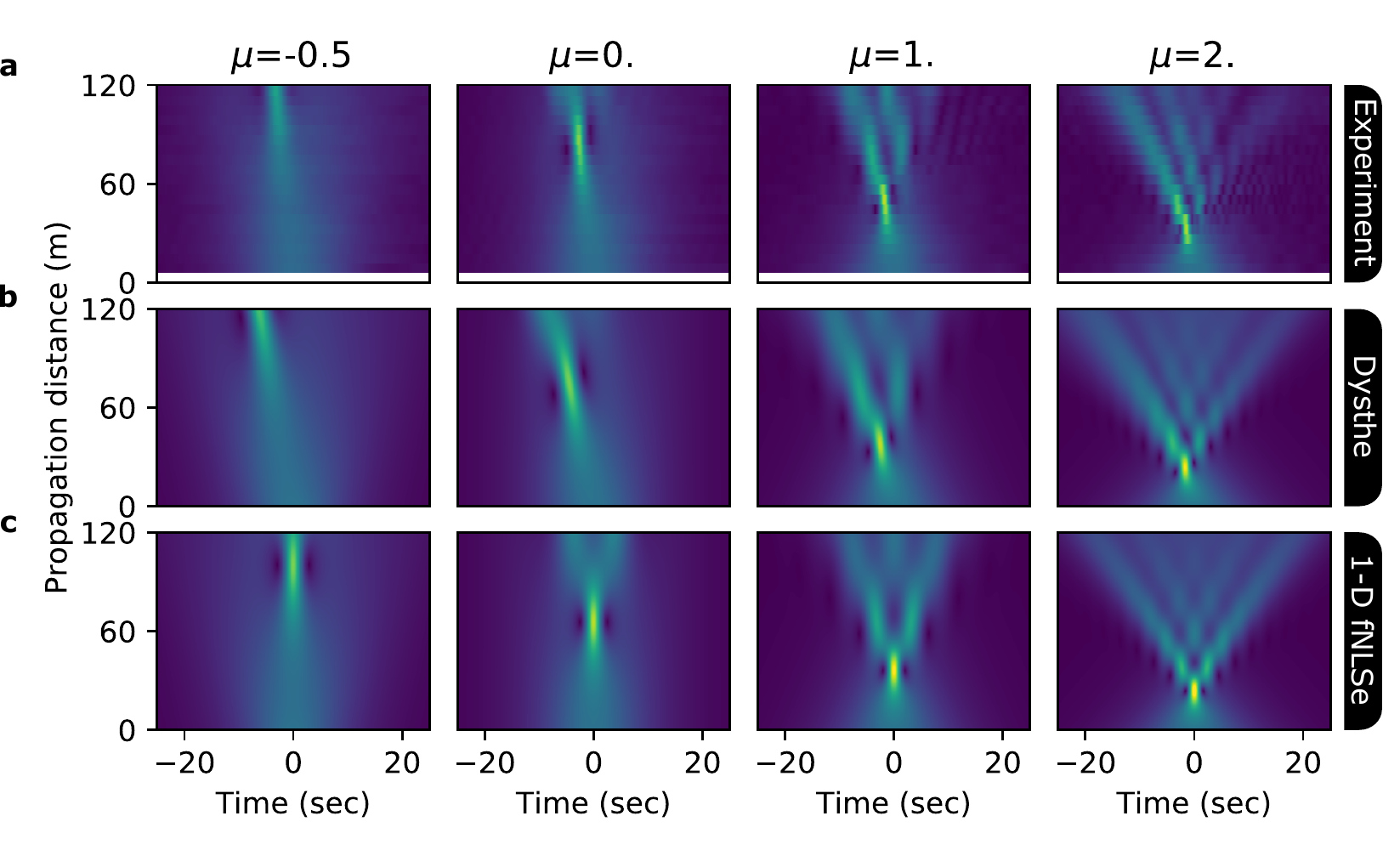}}
\caption{\textbf{ Spatiotemporal evolution. Comparison of the experimental data with numerical simulations of exact 3 soliton solution in Dysthe and NLS models.} \textbf{(a)} spatiotemporal diagram retrieved from the experimental measurements using the Hilbert transform. \textbf{(b)} Simulations of Dysthe equation starting. \textbf{(c)} Simulations of 1-D fNLSe. Exact analytical solution is taken as initial data for the numerical simulations.}
\label{fig:cmp}
\end{figure*}
 
 \begin{figure*}
\center{\includegraphics[width=0.7\linewidth]{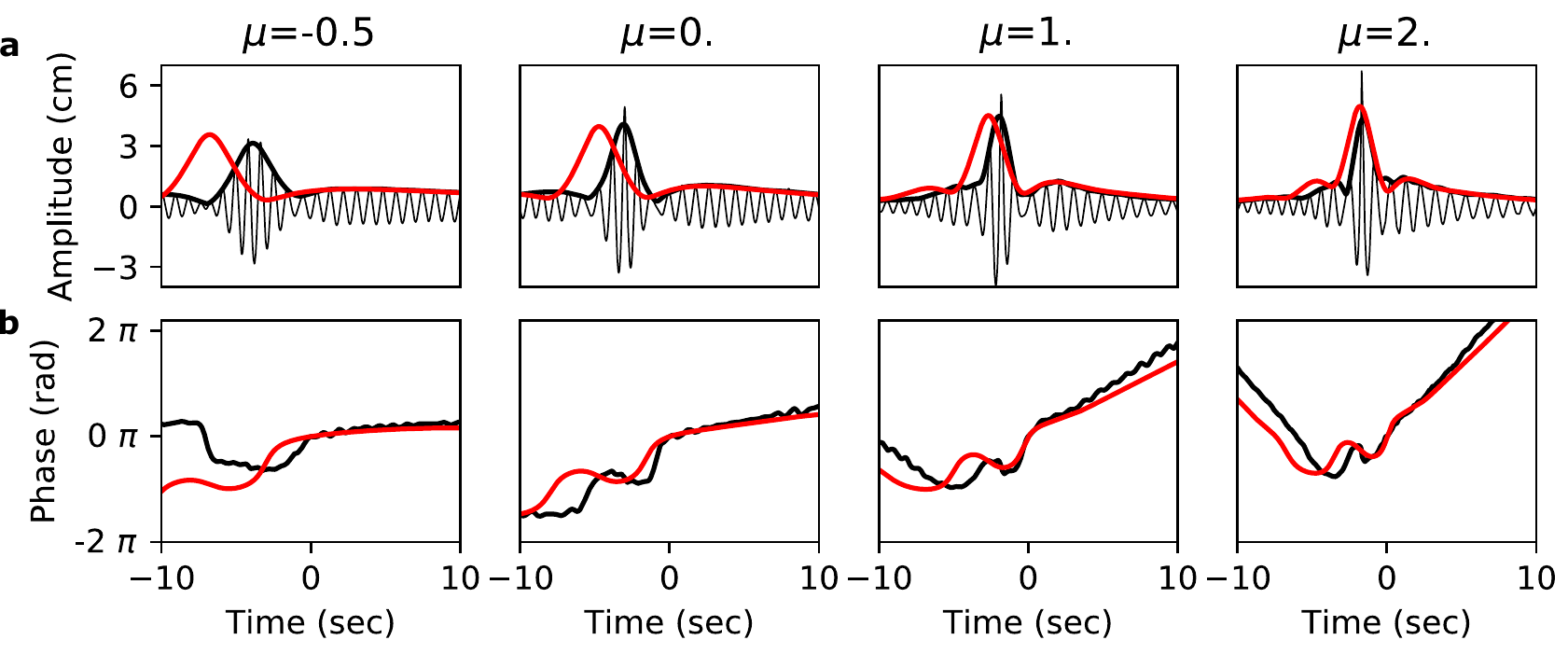}}
\caption{\textbf{Cross-section of Fig.~\ref{fig:cmp} at the maximum compression points.} \textbf{(a)} Measured envelope of water waves (bold black) with underlying carrying wave (black) and simulations of the Dysthe equation (red). \textbf{(b)} Corresponding phase. The positions of maximum compression can be different for numerical simulations and experiment. From left to right the values of $\mu$ are -0.5, 0, 1 and 2.}
\label{fig:ps_emm_exp}
\end{figure*}

\subsubsection{Initial conditions}
We first verify the applicability of the semi-classical theory to the conditions of our water wave experiment. We generate a sequence of wave packets having the same (3 solitons) envelope but different values of the chirp parameter $\mu$ according to Eq.~\eqref{eq:sltlss_ic}. The parameter $\mu$ is varied from -0.8 (the value corresponding to the PS emergence point way beyond the water tank length) to 2 (completely solitonless case). Examples of the initial state measured at the 1st probe (6 m from the wave maker) are plotted in Fig.~\ref{fig:cmp_cut_ic}(a) (black line for the wave elevation). Bold black line shows the calculated envelope of the wave packet by using the Hilbert transform (see Methods section).
Four columns of Fig.~\ref{fig:cmp_cut_ic} correspond to four values of $\mu$: -0.5, 0, 1, and 2. Red line shows the numerical simulations of the Dysthe equation at 6 m distance started from purely numerical initial conditions. Fig.~\ref{fig:cmp_cut_ic}(b) depicts corresponding phase profiles.

The precise control over the IST spectrum is essential for the experimental verification of the result of BT~\cite{Bertola2013Universality}. We analyzed the solitonic content of the initial conditions using the Fourier-collocation method [see Fig.~\ref{fig:cmp_cut_ic} (c)]. As proposed in~\cite{ryczkowski2018real,Chekhovskoy2019Nonlinear}, we solve here the direct stuttering problem and use the discrete part of the IST spectrum for characterizing the solitonic content of the structures under investigation. 
The discrete IST spectra of the water waves are depicted by black crosses, the numerical data at $z=0$~m by blue dots and at $z=6$~m by red dots. We consider the initial conditions having exactly three solitons if the parameter $\mu$ is set to zero as can be seen from the Fig.~\ref{fig:cmp_cut_ic}(c), second column. According to the theoretical prediction~\cite{tovbis_eigenvalue_2000,Bertola2013Universality}, the solitonic content of the pulse changes when the value of the chirp parameter $\mu$ is being varied. In particular, when $\mu$ increases the imaginary part of the discrete eigenvalues decreases until the point when the initial conditions become completely solitonless. This occurs for values of $\mu \ge 2$. Also, from the IST spectra shown in Fig.~\ref{fig:cmp_cut_ic} it follows that the nonlinear propagation over $z=6$~m does not exhibit significant deviation from the integrable case. However, the IST spectra corresponding to $\mu = 2$ already show a deviation from isospectrality due to the presence of the higher-order nonlinear terms in the governing non-integrable (Dysthe) equation.

\subsubsection{Nonlinear evolution}

Nonlinear wave evolution from the initial conditions discussed in the previous section is shown in spatiotemporal diagrams of $|A(z,t)|$, where $A$ is the complex wave envelope (see Fig.~\ref{fig:cmp}). We compare three cases: water tank experiments [plots (a)], simulations of the Dysthe equations [plots (b)] and simulations of the 1-D fNLSe [plots (c)] for four values of the chirp parameter $\mu = -0.5, 0, 1$ and $2$. The experimental data confirms that the maximum compression point can be easily manipulated by adjusting the value of $\mu$. The first column corresponding to $\mu = -0.5$, shows a smooth evolution of the wave packet resulting in the emergence of a localized structure near the edge of the water tank. By increasing the chirp parameter, we observe that the PS maximum compression occurs closer to the wavemaker position while the temporal profile of the local PS packet loses its initial symmetry. In the spatiotemporal diagram corresponding to $\mu = 1$, the coherent structures following the maximum compression point (propagation distance $\approx 72$~m) resemble the localized spikes described in \cite{Bertola2013Universality} despite the broken symmetry. In the solitonless case ($\mu = 2$), the compression point is found as close as $30$~m from the wavemaker, and the subsequent spatiotemporal evolution is confined to  an asymmetric cone.

The experimental data is found to be in an outstanding agreement with numerical simulations of the Dysthe equation [see Fig.~\ref{fig:cmp}(a) and (b)]. The simulations capture the entire spatiotemporal behaviour including small nuances related to the effect of the higher-order nonlinear terms. In order to take into account the effect of finite depth, we include a corresponding coefficient in the simulated model (see Methods section and Appendix in~\cite{Bonnefoy2020modulational}). The evolution of both experimental and simulated signals exhibits the dispersive regularization of gradient catastrophe by the \emph{emergence of a `tilted' local PS-like structure} which follows from the similarity with the corresponding 1-D fNLS model dynamics [Fig.~\ref{fig:cmp}(c)] where the presence of the local PS is a verified fact~\cite{Tikan2017Universality,Tikan2018Single}. The asymmetric shape of the emerging PS-like structure, which has been observed in different hydrodynamic models~\cite{clamond2006long}, is related to the local spectral red-shift which is similar to the effect of Raman scattering in nonlinear fiber optics where the asymmetry of the local PSs has been observed as well, usually in the context of supercontinuum generation~\cite{Dudley2006Supercontinuum}. This also follows from the plots of cross-section at the maximum compression point depicted in Fig.~\ref{fig:ps_emm_exp}. The envelopes of both experimental and simulated signals resemble an asymmetric local PS profile. By increasing the chirp parameter $\mu$, we observed further deviation form the PS shape due to the increasing role of the amplitude-sensitive higher-order nonlinear effects. Indeed, higher values of $\mu$ correspond to the increased pulse amplitude at the gradient catastrophe point, which plays a role of the effective PS background~\cite{Bertola2013Universality,Tikan2017Universality}. The amplitude of the PS (according to Eq.~\eqref{eq:peregrine}) is exactly three times the background value.

\begin{figure}
\center{\includegraphics[width=0.99\linewidth]{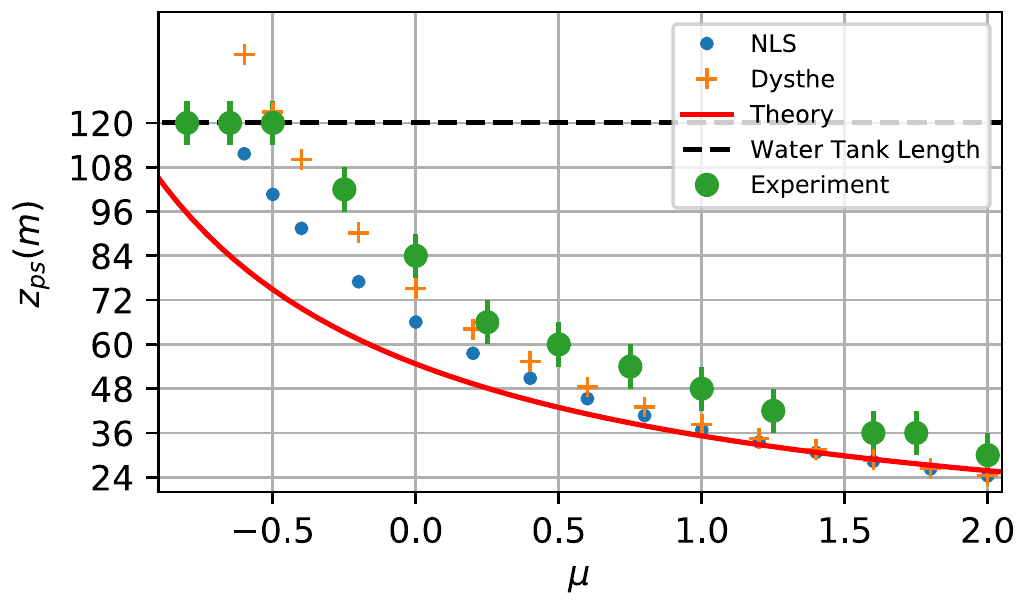}}
\caption{\textbf{Dependence of the Peregrine soliton emergence distance on the initial chirp $\mu$.} Each horizontal gray line represents the position of the probe in the water tank, the black dashed line shows the maximum propagation distance available in the experiments. The positions of the maximum compression point measured experimentally are shown by the green dots with the error bar which corresponds to the distance between probes. It is compared with the simulation of the exact 3 solitons solution in the 1-D fNLSe (blue dots) and the Dysthe equation (orange dots). The prediction according to Eq.~(\ref{eq:PScompressionPoint}) is plotted with the solid red line.}
\label{fig:ps_emm_chart}
\end{figure}

We provide a systematic comparison of the dependence of the local PS-like structures emergence position on the value of the chirp parameter $\mu$ (solid line Fig.~\ref{fig:ps_emm_chart}). A comparison between the data obtained in experiments, simulations of both 1-D fNLS and Dysthe equation, and prediction form the semi-classical theory for $\epsilon$ = 1/3 are shown in Fig.~\ref{fig:ps_emm_chart}. Semi-classical analysis, despite being derived in the zero-dispersion limit, predicts the result of experiments for high values of $\mu$. Moreover, all the data presented in Fig.~\ref{fig:ps_emm_chart} demonstrates the same tendency.
 As followed from spatiotemporal diagrams (Fig.~\ref{fig:cmp}), numerical simulations of the Dysthe equation capture the water wave dynamics. Indeed, corresponding data depicted by orange crosses and greed dots (green ticks indicate a minimum error value related to the distance between the probes) respectively, demonstrate a good agreement along all the values of $\mu$. Experimental data at values of $\mu<-0.5$ are plotted at $z = 120$ m which indicates that the compression point occurs at distances larger than the length of the water tank.

\subsection{IST spectra evolution}

When the 1-D fNLSe is employed for the simulation of nonlinear wave propagation, the global IST spectrum is preserved. However, following the results of the comparison with the experimental data (Fig.~\ref{fig:cmp}), it becomes clear that the higher-order nonlinear effects must be taken into account in order to fully capture the complex self-focusing dynamics affected by the local spectral downshift. The Dysthe equation does not belong to the class of equations integrable by the IST technique. Nonetheless, having an integrable core in the form of 1-D fNLSe, this model can be analysed using perturbation methods if the nonintegrable part is included with a small factor~\cite{Kivshar1989Dynamics}.

Recently, an approach which does not require the higher-order terms to be small has been introduced. In~\cite{ryczkowski2018real,Chekhovskoy2019Nonlinear,sugavanam2019analysis} and~\cite{turitsyn2020nonlinear} it has been shown that IST spectra can be utilized for the characterization of coherent structures in strongly dissipative nonlinear models such as the Ginzburg-Landau equation for mode-locked lasers or the Lugiato-Lefever equation for passive microresonators. The essence of this approach relies on the fact that for each moment of time \emph{the direct scattering problem} --the first step in the IST method~\cite{novikov1984theory} -- can be solved for the given complex envelope of the field which has to be renormalized according to the employed convention~\cite{sugavanam2019analysis}. Therefore, coherent structures that emerge in these complex models can be characterized by a discrete part of the IST spectrum and, therefore, represented by a point in the IST plane. 

\begin{figure}
\center{\includegraphics[width=0.99\linewidth]{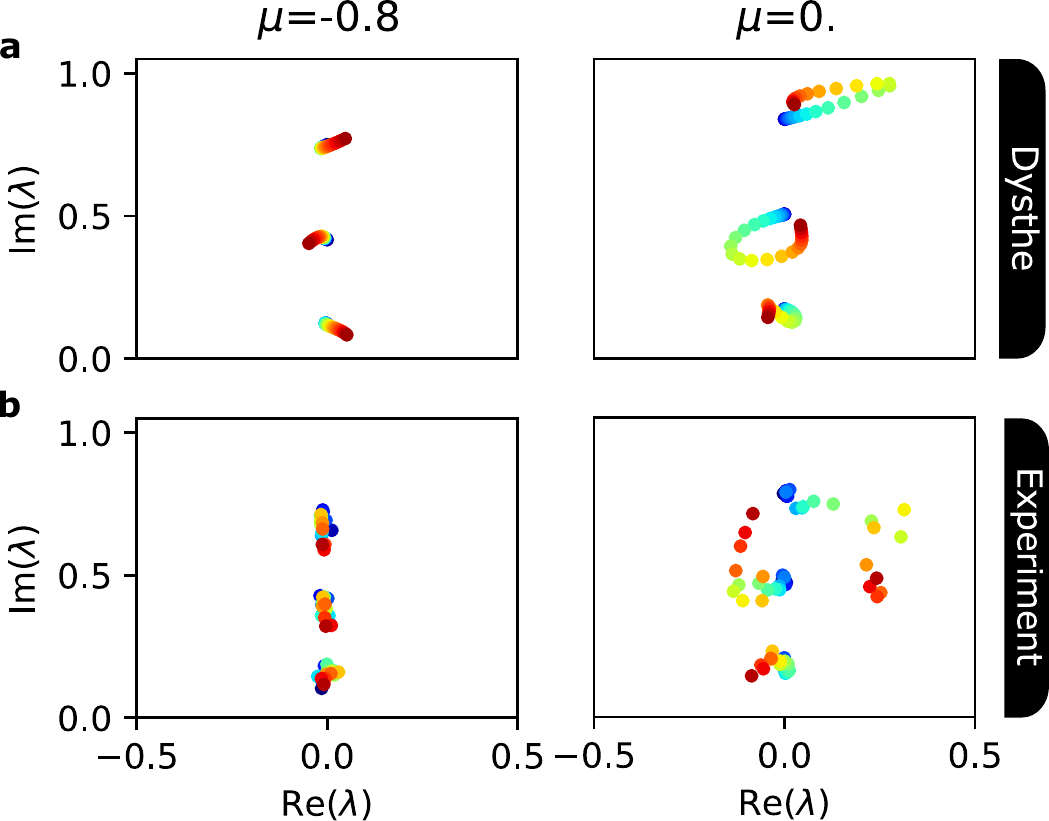}}
\caption{\textbf{Comparison of the discrete IST spectra evolution in the Dysthe model and experiments.} \textbf{(a)} Simulations of the Dysthe equation. \textbf{(b)} Experimental data. Colors from blue to red correspond to the propagation over the full water tank length.}
\label{fig:ist_cmp}
\end{figure}

We use this approach for the characterization of the experimental data and simulations of the Dysthe equation presented in the previous section. For the demonstration, we choose two values of the chirp parameter $\mu$=-0.8 and 0. The resulting IST spectra evolution is shown in Fig.~\ref{fig:ist_cmp}. Color gradient represents the length of the nonlinear propagation and varies from blue ($z=6$~m) to red ($z=120$~m). When the chirp parameter is small the discrete IST spectrum is almost conserved for both the experiments and the Dysthe model~\cite{Suret2020Nonlinear}, which signifies that the nonlinear evolution is close to the one described by the 1-D fNLSe. Indeed, for $\mu = -0.8$ the gradient catastrophe occurs beyond the water tank length so the intensity-related higher-order nonlinear effects can be neglected. The spectral evolution corresponding to the initial condition with $\mu = 0$---the exact three-soliton solution---demonstrates an intriguing behaviour. Namely, the two upper points of the IST spectrum make closed and open loops for simulations and experiment, respectively. The term `closed loop' signifies that the discrete eigenvalues return to their initial positions while the open loop describes the situation when two eigenvalues \textit{exchange positions} during the nonlinear evolution. The presence of a loop signifies that passing through the local PS stage, where the influence of the higher-order nonlinear effects is significant, discrete eigenvalues deviate from their initial position, however, entering the intermediate stage they return to the neighborhood of the origin, thereby, converging to the solution given by the 1-D fNLSe. Investigation of the difference between open and closed loops, as well as a possible exchange of the eigenvalues with the continuum, is beyond of the scope of this manuscript and is proposed as a novel and open problem in the nonintegrable systems evolution.

\section{Discussion}
There are four key conclusions of this article that we would like to highlight:
(i) the local emergence of the PSs as a regularization of the gradient catastrophe can occur in a completely solitonless case, which means that the soliton self-compression is a particular case of this more general process; (ii) the point of the PS emergence can be predicted and easily manipulated by adjusting the chirp parameter, i.e. by controlling the solitonic content of the initial conditions; (iii) structures similar to the local PS are observed in water tank experiments and simulations of the Dysthe model where coherent structures undergo significant red-shift in the spectral domain due to the influence of higher-order terms, which indicates another degree of universality of the gradient catastrophe regularization process; (iv) the IST spectra analysis of the non-integrable nonlinear wave evolution provides insights into the deviation and the convergence to the integrable dynamics as well as reveals unusual behaviour of the discrete points on the IST plane. 

Our experimental observations also suggest that the dynamics of the $N$-soliton solutions rapidly deviates from the one predicted by the 1-D NLS model mainly due to the asymmetry induced by the frequency downshift. However, the local emergence of the coherent structures that can be seen as a modified analogue of the PS can be clearly observed. We, therefore, believe that further extension of the approaches used for the integrable systems to the non-integrable ones can improve our understanding of the realistic mechanisms taking place in the open sea.

\section{Methods}
\subsection{Water tank parameters and mode of operation}

Experimental investigations of the gradient catastrophe regularization for solitonic and solitonless initial conditions have been performed in the water tank facility of the Hydrodynamics, Energetics, and Atmospheric Environment Lab (LHEEA) in Ecole Centrale de Nantes (France).

The water tank is 148~m long (120~m are effectively used), 5~m wide, and 3~m deep. It is equipped with a 8~m long absorbing beach having a parabolic shape. With the addition of pool lanes arranged in a ``W" pattern in front of the beach the measured amplitude reflection coefficient is as low as 1\%. Unidirectional waves are generated with a flap-type wave maker programmed remotely with a computer. The setup comprises of 20 equally-spaced resistive wave-gauges that are installed along the water tank at distances $z_j = 6 + ( j -1)6$ m, where $ j = 1, 2, ...20$ from the wave maker located at $z=0$~m. This provides an effective measuring range of $114$~m.

\subsection{Models and numerical simulations}
1-D fNLSe in the hydrodynamic formulation can be expressed as follows~\cite{Koussaifi2017Spontaneous}:
\begin{equation}
\frac{\partial A}{\partial z} + \frac{i}{g}\frac{\partial^2 A}{\partial t^2} +i\alpha k_0^3 |A|^2A = 0,
\label{eq:NLS}
\end{equation}
here $A$ is the complex envelope of the water wave, $g$ is the acceleration of gravity and $k_0$ - the wave number and $\alpha$ = 0.93 is the finite depth correction. The general expression for $\alpha$ (see e.g.~\cite{mei1989applied} for the details) is stated as:
\begin{equation}\label{gamma}
 \begin{split}
 \alpha= \frac{\cosh{(4k_0 h)}+8-2\tanh^2{(k_0h)}}{8\sinh^4{(k_0 h)}} \\
 - \frac{(2\cosh^2{(k_0 h)}+0.5)^2}{\sinh^2{(2k_0 h)}} \left(\frac{k_0h}{\tanh{(k_0h)}}-\frac{1}{4} \right),
 \end{split}
\end{equation}
where $h$ is the water tank depth.

The Dysthe equation (a higher-order nonlinear generalized version of the 1-D fNLSe) is written in the following way~\cite{goullet2011numerical}:
\begin{widetext}
\begin{equation}
\frac{\partial A}{\partial z} + \frac{i}{g}\frac{\partial^2 A}{\partial t^2} + i \alpha k_0^3 |A|^2A - \frac{k_0^3}{\omega_0} \left[ 6|A|^2\frac{\partial A}{\partial t} + 2A\frac{\partial |A|^2}{\partial t}- iA \mathcal{H}\left( \frac{\partial |A|^2}{\partial t} \right) \right] = 0,
\label{eq:Dysthe}
\end{equation}
\end{widetext}
where $\mathcal{H}$ stands for the Hilbert transform defined as follows:
\[
\mathcal{F}(\mathcal{H}(f(t))) =  -i\,\mathrm{sign}(\omega)\mathcal{F}(f(t)),
\]
where $\mathcal{F}$ represents the Fourier transform and $\mathrm{sign}$ is the signum function .

For the numerical simulations, we used 2048 points simulation box to avoid numerical errors appearing due to the periodicity in the Fourier space. z-axis has been discredited with 1000 point. For simulations of the nonlinear Eq.~\eqref{eq:NLS}and~\eqref{eq:Dysthe}, we use step-adaptive Dormand-Prince Runge-Kutta method of Order 8(5,3) and approximate the dispersion operator by a pseudo-spectral scheme.

\subsection{Calculation of IST spectra}
Introducing the change of variables $\xi = 2\epsilon t$ and $\tau = \epsilon x$ in Eq.~\eqref{eq:NLSeps}, we obtain 1-D NLSe in the following form:
\begin{equation}
\label{eq:NLSmath}
i \frac{\partial u}{\partial t} + \frac{\partial^{2} u}{\partial x^{2}} + 2|u|^{2}u= 0
\end{equation}

We can define a so-called Lax pair for the 1-D fNLSe discovered by Zakharov and Shabat~\cite{Zakharov:1972Exact}:

\begin{equation}
\label{eq:ZSx}
Y_x =
\begin{bmatrix} 
-i \xi & u \\
-u^{*}&i \xi 
\end{bmatrix} Y
\end{equation}

\begin{equation}
\label{eq:ZSt}
Y_t =
\begin{bmatrix} 
-i 2 \xi^{2} + i |u|^{2} & i u_x + 2\xi u \\
i u^{*}_x - 2\xi u^{*} & i 2 \xi^{2} - i |u|^{2} 
\end{bmatrix} Y,
\end{equation}
 where $\xi$ is the spectral parameter and $Y$ is a vector or matrix function.
 Equation~\eqref{eq:NLSmath} is a compatibility condition for Eqs.~\eqref{eq:ZSt} and~\eqref{eq:ZSx} (guaranties the equality of $Y_{tx}$ and $Y_{xt}$).

Equation~\eqref{eq:ZSx} can be inverted to show the spectral problem in more explicit way~\cite{Yang:2010}: 
\begin{equation}
\label{eq:ZSxInv}
\begin{bmatrix} 
- \partial_x & u \\
u^{*}& \partial_x 
\end{bmatrix} Y
= i \xi Y
\end{equation}
This problem can be numerically solved in the Fourier space with a standard routine integrated into the \href{https://docs.scipy.org/doc/scipy/reference/generated/scipy.linalg.eigvals.html#scipy.linalg.eigvals}{SciPy package} of Python.

Numerically retrieved IST spectra are post-processed in order to eliminate points not constituting the discrete part of the spectrum, i.e. not representing the solitonic content of the analysed signal. This is realized by choosing a threshold on the imaginary axis of the IST spectrum (eigenvalues of the spectral problem represented by Eq.~\eqref{eq:ZSxInv}) and keeping only the eigenvalues that exceed the value. The threshold can be found empirically by changing the box discretization which affects non-representative part of the spectrum while keeping the actual IST spectrum unchanged within a reasonable range of parameters.

\section{Acknowledgments}
This work has been partially supported by the Agence Nationale de la
Recherche through the LABEX CEMPI project (ANR-11-LABX-0007) and the ANR DYSTURB Project (ANR-17-CE30-0004),  the
Ministry of Higher Education and Research, Hauts de France council and
European Regional Development Fund (ERDF) through the Nord-Pas de
Calais Regional Research Council and the European Regional Development
Fund (ERDF) through the Contrat de Projets Etat-R\'egion (CPER
Photonics for Society P4S). E.F. is thankful for partial support from the Simons Foundation/MPS No. 651463.  The work was also partially supported by the EPSRC (UK) grant 
EP/R00515X/2 (G.E.), NSF (USA) grant DMS 2009647 (A.T.) and Dstl (UK) grant DSTLX-1000116851 (G.R., G.E., S.R.).


\bibliography{Peregrine_hydro}

\begin{thebibliography}{55}%
\makeatletter
\providecommand \@ifxundefined [1]{%
 \@ifx{#1\undefined}
}%
\providecommand \@ifnum [1]{%
 \ifnum #1\expandafter \@firstoftwo
 \else \expandafter \@secondoftwo
 \fi
}%
\providecommand \@ifx [1]{%
 \ifx #1\expandafter \@firstoftwo
 \else \expandafter \@secondoftwo
 \fi
}%
\providecommand \natexlab [1]{#1}%
\providecommand \enquote  [1]{``#1''}%
\providecommand \bibnamefont  [1]{#1}%
\providecommand \bibfnamefont [1]{#1}%
\providecommand \citenamefont [1]{#1}%
\providecommand \href@noop [0]{\@secondoftwo}%
\providecommand \href [0]{\begingroup \@sanitize@url \@href}%
\providecommand \@href[1]{\@@startlink{#1}\@@href}%
\providecommand \@@href[1]{\endgroup#1\@@endlink}%
\providecommand \@sanitize@url [0]{\catcode `\\12\catcode `\$12\catcode
  `\&12\catcode `\#12\catcode `\^12\catcode `\_12\catcode `\%12\relax}%
\providecommand \@@startlink[1]{}%
\providecommand \@@endlink[0]{}%
\providecommand \url  [0]{\begingroup\@sanitize@url \@url }%
\providecommand \@url [1]{\endgroup\@href {#1}{\urlprefix }}%
\providecommand \urlprefix  [0]{URL }%
\providecommand \Eprint [0]{\href }%
\providecommand \doibase [0]{http://dx.doi.org/}%
\providecommand \selectlanguage [0]{\@gobble}%
\providecommand \bibinfo  [0]{\@secondoftwo}%
\providecommand \bibfield  [0]{\@secondoftwo}%
\providecommand \translation [1]{[#1]}%
\providecommand \BibitemOpen [0]{}%
\providecommand \bibitemStop [0]{}%
\providecommand \bibitemNoStop [0]{.\EOS\space}%
\providecommand \EOS [0]{\spacefactor3000\relax}%
\providecommand \BibitemShut  [1]{\csname bibitem#1\endcsname}%
\let\auto@bib@innerbib\@empty
\bibitem [{\citenamefont {Annenkov}\ and\ \citenamefont
  {Shrira}(2001)}]{Annenkov2001predictability}%
  \BibitemOpen
  \bibfield  {author} {\bibinfo {author} {\bibfnamefont {S.}~\bibnamefont
  {Annenkov}}\ and\ \bibinfo {author} {\bibfnamefont {V.~I.}\ \bibnamefont
  {Shrira}},\ }\href {\doibase 10.1016/S0167-2789(01)00199-3} {\bibfield
  {journal} {\bibinfo  {journal} {Physica D}\ }\textbf {\bibinfo {volume}
  {152-153}},\ \bibinfo {pages} {665} (\bibinfo {year} {2001})}\BibitemShut
  {NoStop}%
\bibitem [{\citenamefont {Kharif}\ \emph {et~al.}(2008)\citenamefont {Kharif},
  \citenamefont {Pelinovsky},\ and\ \citenamefont
  {Slunyaev}}]{kharif2008Rogue}%
  \BibitemOpen
  \bibfield  {author} {\bibinfo {author} {\bibfnamefont {C.}~\bibnamefont
  {Kharif}}, \bibinfo {author} {\bibfnamefont {E.}~\bibnamefont {Pelinovsky}},
  \ and\ \bibinfo {author} {\bibfnamefont {A.}~\bibnamefont {Slunyaev}},\
  }\href@noop {} {\emph {\bibinfo {title} {Rogue waves in the ocean}}}\
  (\bibinfo  {publisher} {Springer Science \& Business Media},\ \bibinfo {year}
  {2008})\BibitemShut {NoStop}%
\bibitem [{\citenamefont {Onorato}\ \emph {et~al.}(2013)\citenamefont
  {Onorato}, \citenamefont {Residori}, \citenamefont {Bortolozzo},
  \citenamefont {Montina},\ and\ \citenamefont {Arecchi}}]{Onorato2013Rogue}%
  \BibitemOpen
  \bibfield  {author} {\bibinfo {author} {\bibfnamefont {M.}~\bibnamefont
  {Onorato}}, \bibinfo {author} {\bibfnamefont {S.}~\bibnamefont {Residori}},
  \bibinfo {author} {\bibfnamefont {U.}~\bibnamefont {Bortolozzo}}, \bibinfo
  {author} {\bibfnamefont {A.}~\bibnamefont {Montina}}, \ and\ \bibinfo
  {author} {\bibfnamefont {F.}~\bibnamefont {Arecchi}},\ }\href@noop {}
  {\bibfield  {journal} {\bibinfo  {journal} {Physics Reports}\ }\textbf
  {\bibinfo {volume} {528}},\ \bibinfo {pages} {47} (\bibinfo {year}
  {2013})}\BibitemShut {NoStop}%
\bibitem [{\citenamefont {Cousins}\ and\ \citenamefont
  {Sapsis}(2014)}]{Cousins2014Quantification}%
  \BibitemOpen
  \bibfield  {author} {\bibinfo {author} {\bibfnamefont {W.}~\bibnamefont
  {Cousins}}\ and\ \bibinfo {author} {\bibfnamefont {T.~P.}\ \bibnamefont
  {Sapsis}},\ }\href {\doibase 10.1016/j.physd.2014.04.012} {\bibfield
  {journal} {\bibinfo  {journal} {Physica D}\ }\textbf {\bibinfo {volume}
  {280-281}},\ \bibinfo {pages} {48} (\bibinfo {year} {2014})}\BibitemShut
  {NoStop}%
\bibitem [{\citenamefont {Birkholz}\ \emph {et~al.}(2015)\citenamefont
  {Birkholz}, \citenamefont {Br\'ee}, \citenamefont {Demircan},\ and\
  \citenamefont {Steinmeyer}}]{Birkholz2015Predictability}%
  \BibitemOpen
  \bibfield  {author} {\bibinfo {author} {\bibfnamefont {S.}~\bibnamefont
  {Birkholz}}, \bibinfo {author} {\bibfnamefont {C.}~\bibnamefont {Br\'ee}},
  \bibinfo {author} {\bibfnamefont {A.}~\bibnamefont {Demircan}}, \ and\
  \bibinfo {author} {\bibfnamefont {G.}~\bibnamefont {Steinmeyer}},\ }\href
  {\doibase 10.1103/PhysRevLett.114.213901} {\bibfield  {journal} {\bibinfo
  {journal} {Phys. Rev. Lett.}\ }\textbf {\bibinfo {volume} {114}},\ \bibinfo
  {pages} {213901} (\bibinfo {year} {2015})}\BibitemShut {NoStop}%
\bibitem [{\citenamefont {Cousins}\ \emph {et~al.}(2019)\citenamefont
  {Cousins}, \citenamefont {Onorato}, \citenamefont {Chabchoub},\ and\
  \citenamefont {Sapsis}}]{Cousins2019Predicting}%
  \BibitemOpen
  \bibfield  {author} {\bibinfo {author} {\bibfnamefont {W.}~\bibnamefont
  {Cousins}}, \bibinfo {author} {\bibfnamefont {M.}~\bibnamefont {Onorato}},
  \bibinfo {author} {\bibfnamefont {A.}~\bibnamefont {Chabchoub}}, \ and\
  \bibinfo {author} {\bibfnamefont {T.~P.}\ \bibnamefont {Sapsis}},\ }\href
  {\doibase 10.1103/PhysRevE.99.032201} {\bibfield  {journal} {\bibinfo
  {journal} {Phys. Rev. E}\ }\textbf {\bibinfo {volume} {99}},\ \bibinfo
  {pages} {032201} (\bibinfo {year} {2019})}\BibitemShut {NoStop}%
\bibitem [{\citenamefont {Dudley}\ \emph {et~al.}(2019)\citenamefont {Dudley},
  \citenamefont {Genty}, \citenamefont {Mussot}, \citenamefont {Chabchoub},\
  and\ \citenamefont {Dias}}]{dudley2019rogue}%
  \BibitemOpen
  \bibfield  {author} {\bibinfo {author} {\bibfnamefont {J.~M.}\ \bibnamefont
  {Dudley}}, \bibinfo {author} {\bibfnamefont {G.}~\bibnamefont {Genty}},
  \bibinfo {author} {\bibfnamefont {A.}~\bibnamefont {Mussot}}, \bibinfo
  {author} {\bibfnamefont {A.}~\bibnamefont {Chabchoub}}, \ and\ \bibinfo
  {author} {\bibfnamefont {F.}~\bibnamefont {Dias}},\ }\href@noop {} {\bibfield
   {journal} {\bibinfo  {journal} {Nature Reviews Physics}\ }\textbf {\bibinfo
  {volume} {1}},\ \bibinfo {pages} {675} (\bibinfo {year} {2019})}\BibitemShut
  {NoStop}%
\bibitem [{\citenamefont {Onorato}\ \emph {et~al.}(2004)\citenamefont
  {Onorato}, \citenamefont {Osborne}, \citenamefont {Serio}, \citenamefont
  {Cavaleri}, \citenamefont {Brandini},\ and\ \citenamefont
  {Stansberg}}]{Onorato2004Observation}%
  \BibitemOpen
  \bibfield  {author} {\bibinfo {author} {\bibfnamefont {M.}~\bibnamefont
  {Onorato}}, \bibinfo {author} {\bibfnamefont {A.~R.}\ \bibnamefont
  {Osborne}}, \bibinfo {author} {\bibfnamefont {M.}~\bibnamefont {Serio}},
  \bibinfo {author} {\bibfnamefont {L.}~\bibnamefont {Cavaleri}}, \bibinfo
  {author} {\bibfnamefont {C.}~\bibnamefont {Brandini}}, \ and\ \bibinfo
  {author} {\bibfnamefont {C.~T.}\ \bibnamefont {Stansberg}},\ }\href {\doibase
  10.1103/PhysRevE.70.067302} {\bibfield  {journal} {\bibinfo  {journal} {Phys.
  Rev. E}\ }\textbf {\bibinfo {volume} {70}},\ \bibinfo {pages} {067302}
  (\bibinfo {year} {2004})}\BibitemShut {NoStop}%
\bibitem [{\citenamefont {Wabnitz}\ \emph {et~al.}(2016)\citenamefont
  {Wabnitz}, \citenamefont {Reid}, \citenamefont {Heyl}, \citenamefont
  {Thomson}, \citenamefont {Akhmediev}, \citenamefont {Kibler}, \citenamefont
  {Baronio},\ and\ \citenamefont {Beli}}]{Wabnitz2016Roadmap}%
  \BibitemOpen
  \bibfield  {author} {\bibinfo {author} {\bibfnamefont {S.}~\bibnamefont
  {Wabnitz}}, \bibinfo {author} {\bibfnamefont {D.~T.}\ \bibnamefont {Reid}},
  \bibinfo {author} {\bibfnamefont {C.~M.}\ \bibnamefont {Heyl}}, \bibinfo
  {author} {\bibfnamefont {R.~R.}\ \bibnamefont {Thomson}}, \bibinfo {author}
  {\bibfnamefont {N.}~\bibnamefont {Akhmediev}}, \bibinfo {author}
  {\bibfnamefont {B.}~\bibnamefont {Kibler}}, \bibinfo {author} {\bibfnamefont
  {F.}~\bibnamefont {Baronio}}, \ and\ \bibinfo {author} {\bibfnamefont
  {M.}~\bibnamefont {Beli}},\ }\href {\doibase 10.1088/2040-8978/18/6/063001}
  {\bibfield  {journal} {\bibinfo  {journal} {J. Opt.}\ }\textbf {\bibinfo
  {volume} {18}} (\bibinfo {year} {2016}),\
  10.1088/2040-8978/18/6/063001}\BibitemShut {NoStop}%
\bibitem [{\citenamefont {Randoux}\ \emph {et~al.}(2016)\citenamefont
  {Randoux}, \citenamefont {Walczak}, \citenamefont {Onorato},\ and\
  \citenamefont {Suret}}]{Randoux2016Nonlinear}%
  \BibitemOpen
  \bibfield  {author} {\bibinfo {author} {\bibfnamefont {S.}~\bibnamefont
  {Randoux}}, \bibinfo {author} {\bibfnamefont {P.}~\bibnamefont {Walczak}},
  \bibinfo {author} {\bibfnamefont {M.}~\bibnamefont {Onorato}}, \ and\
  \bibinfo {author} {\bibfnamefont {P.}~\bibnamefont {Suret}},\ }\href
  {\doibase 10.1016/j.physd.2016.04.001} {\bibfield  {journal} {\bibinfo
  {journal} {Physica D}\ }\textbf {\bibinfo {volume} {333}},\ \bibinfo {pages}
  {323} (\bibinfo {year} {2016})}\BibitemShut {NoStop}%
\bibitem [{\citenamefont {El~Koussaifi}\ \emph {et~al.}(2018)\citenamefont
  {El~Koussaifi}, \citenamefont {Tikan}, \citenamefont {Toffoli}, \citenamefont
  {Randoux}, \citenamefont {Suret},\ and\ \citenamefont
  {Onorato}}]{Koussaifi2017Spontaneous}%
  \BibitemOpen
  \bibfield  {author} {\bibinfo {author} {\bibfnamefont {R.}~\bibnamefont
  {El~Koussaifi}}, \bibinfo {author} {\bibfnamefont {A.}~\bibnamefont {Tikan}},
  \bibinfo {author} {\bibfnamefont {A.}~\bibnamefont {Toffoli}}, \bibinfo
  {author} {\bibfnamefont {S.}~\bibnamefont {Randoux}}, \bibinfo {author}
  {\bibfnamefont {P.}~\bibnamefont {Suret}}, \ and\ \bibinfo {author}
  {\bibfnamefont {M.}~\bibnamefont {Onorato}},\ }\href {\doibase
  10.1103/PhysRevE.97.012208} {\bibfield  {journal} {\bibinfo  {journal} {Phys.
  Rev. E}\ }\textbf {\bibinfo {volume} {97}},\ \bibinfo {pages} {012208}
  (\bibinfo {year} {2018})}\BibitemShut {NoStop}%
\bibitem [{\citenamefont {Zakharov}\ and\ \citenamefont
  {Shabat}(1972)}]{Zakharov:1972Exact}%
  \BibitemOpen
  \bibfield  {author} {\bibinfo {author} {\bibfnamefont {V.~E.}\ \bibnamefont
  {Zakharov}}\ and\ \bibinfo {author} {\bibfnamefont {A.~B.}\ \bibnamefont
  {Shabat}},\ }\href@noop {} {\bibfield  {journal} {\bibinfo  {journal} {Sov.
  Phys. JETP}\ }\textbf {\bibinfo {volume} {34}},\ \bibinfo {pages} {62}
  (\bibinfo {year} {1972})}\BibitemShut {NoStop}%
\bibitem [{\citenamefont {Zakharov}(1968)}]{Zakharov1968Stability}%
  \BibitemOpen
  \bibfield  {author} {\bibinfo {author} {\bibfnamefont {V.~E.}\ \bibnamefont
  {Zakharov}},\ }\href {\doibase 10.1007/BF00913182} {\bibfield  {journal}
  {\bibinfo  {journal} {J. Appl. Mech. Tech. Phys.}\ }\textbf {\bibinfo
  {volume} {9}},\ \bibinfo {pages} {190} (\bibinfo {year} {1968})}\BibitemShut
  {NoStop}%
\bibitem [{\citenamefont {Novikov}\ \emph {et~al.}(1984)\citenamefont
  {Novikov}, \citenamefont {Manakov}, \citenamefont {Pitaevskii},\ and\
  \citenamefont {Zakharov}}]{novikov1984theory}%
  \BibitemOpen
  \bibfield  {author} {\bibinfo {author} {\bibfnamefont {S.}~\bibnamefont
  {Novikov}}, \bibinfo {author} {\bibfnamefont {S.}~\bibnamefont {Manakov}},
  \bibinfo {author} {\bibfnamefont {L.}~\bibnamefont {Pitaevskii}}, \ and\
  \bibinfo {author} {\bibfnamefont {V.~E.}\ \bibnamefont {Zakharov}},\
  }\href@noop {} {\emph {\bibinfo {title} {Theory of solitons: the inverse
  scattering method}}}\ (\bibinfo  {publisher} {Springer Science \& Business
  Media},\ \bibinfo {year} {1984})\BibitemShut {NoStop}%
\bibitem [{\citenamefont {Zakharov}(2009)}]{Zakharov2009turbulence}%
  \BibitemOpen
  \bibfield  {author} {\bibinfo {author} {\bibfnamefont {V.~E.}\ \bibnamefont
  {Zakharov}},\ }\href {\doibase 10.1111/j.1467-9590.2009.00430.x} {\bibfield
  {journal} {\bibinfo  {journal} {Stud. Appl. Math.}\ }\textbf {\bibinfo
  {volume} {122}},\ \bibinfo {pages} {219} (\bibinfo {year}
  {2009})}\BibitemShut {NoStop}%
\bibitem [{\citenamefont {Soto-Crespo}\ \emph {et~al.}(2016)\citenamefont
  {Soto-Crespo}, \citenamefont {Devine},\ and\ \citenamefont
  {Akhmediev}}]{Soto-Crespo2016Integrable}%
  \BibitemOpen
  \bibfield  {author} {\bibinfo {author} {\bibfnamefont {J.~M.}\ \bibnamefont
  {Soto-Crespo}}, \bibinfo {author} {\bibfnamefont {N.}~\bibnamefont {Devine}},
  \ and\ \bibinfo {author} {\bibfnamefont {N.}~\bibnamefont {Akhmediev}},\
  }\href {\doibase 10.1103/PhysRevLett.116.103901} {\bibfield  {journal}
  {\bibinfo  {journal} {Phys. Rev. Lett.}\ }\textbf {\bibinfo {volume} {116}},\
  \bibinfo {pages} {103901} (\bibinfo {year} {2016})}\BibitemShut {NoStop}%
\bibitem [{\citenamefont {Akhmediev}\ \emph {et~al.}(2016)\citenamefont
  {Akhmediev}, \citenamefont {Soto-Crespo},\ and\ \citenamefont
  {Devine}}]{Akhmediev2016Breather}%
  \BibitemOpen
  \bibfield  {author} {\bibinfo {author} {\bibfnamefont {N.~N.}\ \bibnamefont
  {Akhmediev}}, \bibinfo {author} {\bibfnamefont {J.~M.}\ \bibnamefont
  {Soto-Crespo}}, \ and\ \bibinfo {author} {\bibfnamefont {N.}~\bibnamefont
  {Devine}},\ }\href {\doibase 10.1103/PhysRevE.94.022212} {\bibfield
  {journal} {\bibinfo  {journal} {Phys. Rev. E}\ }\textbf {\bibinfo {volume}
  {94}},\ \bibinfo {pages} {022212} (\bibinfo {year} {2016})}\BibitemShut
  {NoStop}%
\bibitem [{\citenamefont {Agafontsev}\ and\ \citenamefont
  {Zakharov}(2015)}]{Agafontsev2015Integrable}%
  \BibitemOpen
  \bibfield  {author} {\bibinfo {author} {\bibfnamefont {D.~S.}\ \bibnamefont
  {Agafontsev}}\ and\ \bibinfo {author} {\bibfnamefont {V.~E.}\ \bibnamefont
  {Zakharov}},\ }\href {\doibase 10.1088/0951-7715/28/8/2791} {\bibfield
  {journal} {\bibinfo  {journal} {Nonlinearity}\ }\textbf {\bibinfo {volume}
  {28}},\ \bibinfo {pages} {2791} (\bibinfo {year} {2015})}\BibitemShut
  {NoStop}%
\bibitem [{\citenamefont {Kraych}\ \emph {et~al.}(2019)\citenamefont {Kraych},
  \citenamefont {Agafontsev}, \citenamefont {Randoux},\ and\ \citenamefont
  {Suret}}]{kraych2019statistical}%
  \BibitemOpen
  \bibfield  {author} {\bibinfo {author} {\bibfnamefont {A.~E.}\ \bibnamefont
  {Kraych}}, \bibinfo {author} {\bibfnamefont {D.}~\bibnamefont {Agafontsev}},
  \bibinfo {author} {\bibfnamefont {S.}~\bibnamefont {Randoux}}, \ and\
  \bibinfo {author} {\bibfnamefont {P.}~\bibnamefont {Suret}},\ }\href
  {\doibase 10.1103/PhysRevLett.123.093902} {\bibfield  {journal} {\bibinfo
  {journal} {Phys. Rev. Lett.}\ }\textbf {\bibinfo {volume} {123}},\ \bibinfo
  {pages} {093902} (\bibinfo {year} {2019})}\BibitemShut {NoStop}%
\bibitem [{\citenamefont {Copie}\ \emph {et~al.}(2020)\citenamefont {Copie},
  \citenamefont {Randoux},\ and\ \citenamefont {Suret}}]{copie2019physics}%
  \BibitemOpen
  \bibfield  {author} {\bibinfo {author} {\bibfnamefont {F.}~\bibnamefont
  {Copie}}, \bibinfo {author} {\bibfnamefont {S.}~\bibnamefont {Randoux}}, \
  and\ \bibinfo {author} {\bibfnamefont {P.}~\bibnamefont {Suret}},\ }\href
  {\doibase https://doi.org/10.1016/j.revip.2019.100037} {\bibfield  {journal}
  {\bibinfo  {journal} {Reviews in Physics}\ ,\ \bibinfo {pages} {100037}}
  (\bibinfo {year} {2020})}\BibitemShut {NoStop}%
\bibitem [{\citenamefont {Walczak}\ \emph {et~al.}(2015)\citenamefont
  {Walczak}, \citenamefont {Randoux},\ and\ \citenamefont
  {Suret}}]{Walczak2015Optical}%
  \BibitemOpen
  \bibfield  {author} {\bibinfo {author} {\bibfnamefont {P.}~\bibnamefont
  {Walczak}}, \bibinfo {author} {\bibfnamefont {S.}~\bibnamefont {Randoux}}, \
  and\ \bibinfo {author} {\bibfnamefont {P.}~\bibnamefont {Suret}},\ }\href
  {\doibase 10.1103/PhysRevLett.114.143903} {\bibfield  {journal} {\bibinfo
  {journal} {Phys. Rev. Lett.}\ }\textbf {\bibinfo {volume} {114}},\ \bibinfo
  {pages} {143903} (\bibinfo {year} {2015})}\BibitemShut {NoStop}%
\bibitem [{\citenamefont {Suret}\ \emph {et~al.}(2016)\citenamefont {Suret},
  \citenamefont {Koussaifi}, \citenamefont {Tikan}, \citenamefont {Evain},
  \citenamefont {Randoux}, \citenamefont {Szwaj},\ and\ \citenamefont
  {Bielawski}}]{Suret2016Single}%
  \BibitemOpen
  \bibfield  {author} {\bibinfo {author} {\bibfnamefont {P.}~\bibnamefont
  {Suret}}, \bibinfo {author} {\bibfnamefont {R.~E.}\ \bibnamefont
  {Koussaifi}}, \bibinfo {author} {\bibfnamefont {A.}~\bibnamefont {Tikan}},
  \bibinfo {author} {\bibfnamefont {C.}~\bibnamefont {Evain}}, \bibinfo
  {author} {\bibfnamefont {S.}~\bibnamefont {Randoux}}, \bibinfo {author}
  {\bibfnamefont {C.}~\bibnamefont {Szwaj}}, \ and\ \bibinfo {author}
  {\bibfnamefont {S.}~\bibnamefont {Bielawski}},\ }\href {\doibase
  10.1038/ncomms13136} {\bibfield  {journal} {\bibinfo  {journal} {Nat.
  Commun.}\ }\textbf {\bibinfo {volume} {7}},\ \bibinfo {pages} {13136}
  (\bibinfo {year} {2016})}\BibitemShut {NoStop}%
\bibitem [{\citenamefont {Suret}\ \emph {et~al.}(2017)\citenamefont {Suret},
  \citenamefont {El}, \citenamefont {Onorato},\ and\ \citenamefont
  {Randoux}}]{Suret:2017book}%
  \BibitemOpen
  \bibfield  {author} {\bibinfo {author} {\bibfnamefont {P.}~\bibnamefont
  {Suret}}, \bibinfo {author} {\bibfnamefont {G.}~\bibnamefont {El}}, \bibinfo
  {author} {\bibfnamefont {M.}~\bibnamefont {Onorato}}, \ and\ \bibinfo
  {author} {\bibfnamefont {S.}~\bibnamefont {Randoux}},\ }\enquote {\bibinfo
  {title} {Rogue waves in integrable turbulence: semi-classical theory and fast
  measurements},}\ in\ \href {\doibase 10.1088/978-0-7503-1460-2ch12} {\emph
  {\bibinfo {booktitle} {Nonlinear Guided Wave Optics}}},\ \bibinfo {series and
  number} {2053-2563}\ (\bibinfo  {publisher} {IOP Publishing},\ \bibinfo
  {year} {2017})\ pp.\ \bibinfo {pages} {12--1 to 12--32}\BibitemShut {NoStop}%
\bibitem [{\citenamefont {Kuznetsov}(1977)}]{kuznetsov1977solitons}%
  \BibitemOpen
  \bibfield  {author} {\bibinfo {author} {\bibfnamefont {E.~A.}\ \bibnamefont
  {Kuznetsov}},\ }\href@noop {} {\bibfield  {journal} {\bibinfo  {journal}
  {Doklady Akademii Nauk}\ }\textbf {\bibinfo {volume} {236}},\ \bibinfo
  {pages} {575} (\bibinfo {year} {1977})}\BibitemShut {NoStop}%
\bibitem [{\citenamefont {Ma}(1979)}]{Ma1979Perturbed}%
  \BibitemOpen
  \bibfield  {author} {\bibinfo {author} {\bibfnamefont {Y.}~\bibnamefont
  {Ma}},\ }\href {\doibase 10.1002/sapm197960143} {\bibfield  {journal}
  {\bibinfo  {journal} {Stud. Appl. Math}\ }\textbf {\bibinfo {volume} {60}},\
  \bibinfo {pages} {43} (\bibinfo {year} {1979})}\BibitemShut {NoStop}%
\bibitem [{\citenamefont {Peregrine}(1983)}]{peregrine1983water}%
  \BibitemOpen
  \bibfield  {author} {\bibinfo {author} {\bibfnamefont {D.~H.}\ \bibnamefont
  {Peregrine}},\ }\href {\doibase 10.1017/S0334270000003891} {\bibfield
  {journal} {\bibinfo  {journal} {The Journal of the Australian Mathematical
  Society. Series B. Applied Mathematics}\ }\textbf {\bibinfo {volume} {25}},\
  \bibinfo {pages} {16–43} (\bibinfo {year} {1983})}\BibitemShut {NoStop}%
\bibitem [{\citenamefont {Akhmediev}\ and\ \citenamefont
  {Korneev}(1986)}]{akhmediev1986modulation}%
  \BibitemOpen
  \bibfield  {author} {\bibinfo {author} {\bibfnamefont {N.~N.}\ \bibnamefont
  {Akhmediev}}\ and\ \bibinfo {author} {\bibfnamefont {V.~I.}\ \bibnamefont
  {Korneev}},\ }\href {\doibase 10.1007/BF01037866} {\bibfield  {journal}
  {\bibinfo  {journal} {Theor. Math. Phys.}\ }\textbf {\bibinfo {volume}
  {69}},\ \bibinfo {pages} {1089} (\bibinfo {year} {1986})}\BibitemShut
  {NoStop}%
\bibitem [{\citenamefont {Shrira}\ and\ \citenamefont
  {Geogjaev}(2010)}]{Shrira2010What}%
  \BibitemOpen
  \bibfield  {author} {\bibinfo {author} {\bibfnamefont {V.~I.}\ \bibnamefont
  {Shrira}}\ and\ \bibinfo {author} {\bibfnamefont {V.~V.}\ \bibnamefont
  {Geogjaev}},\ }\href {\doibase 10.1007/s10665-009-9347-2} {\bibfield
  {journal} {\bibinfo  {journal} {J. Eng. Math.}\ }\textbf {\bibinfo {volume}
  {67}},\ \bibinfo {pages} {11} (\bibinfo {year} {2010})}\BibitemShut {NoStop}%
\bibitem [{\citenamefont {Dysthe}\ and\ \citenamefont
  {Trulsen}(1999)}]{Dysthe1999Note}%
  \BibitemOpen
  \bibfield  {author} {\bibinfo {author} {\bibfnamefont {K.~B.}\ \bibnamefont
  {Dysthe}}\ and\ \bibinfo {author} {\bibfnamefont {K.}~\bibnamefont
  {Trulsen}},\ }\href {\doibase 10.1238/Physica.Topical.082a00048} {\bibfield
  {journal} {\bibinfo  {journal} {Phys. Scr.}\ }\textbf {\bibinfo {volume}
  {T82}},\ \bibinfo {pages} {48} (\bibinfo {year} {1999})}\BibitemShut
  {NoStop}%
\bibitem [{\citenamefont {Dudley}\ \emph {et~al.}(2014)\citenamefont {Dudley},
  \citenamefont {Dias}, \citenamefont {Erkintalo},\ and\ \citenamefont
  {Genty}}]{Dudley2014Instabilities}%
  \BibitemOpen
  \bibfield  {author} {\bibinfo {author} {\bibfnamefont {J.~M.}\ \bibnamefont
  {Dudley}}, \bibinfo {author} {\bibfnamefont {F.}~\bibnamefont {Dias}},
  \bibinfo {author} {\bibfnamefont {M.}~\bibnamefont {Erkintalo}}, \ and\
  \bibinfo {author} {\bibfnamefont {G.}~\bibnamefont {Genty}},\ }\href
  {\doibase 10.1038/nphoton.2014.220} {\bibfield  {journal} {\bibinfo
  {journal} {Nat. Photonics}\ }\textbf {\bibinfo {volume} {8}},\ \bibinfo
  {pages} {755} (\bibinfo {year} {2014})}\BibitemShut {NoStop}%
\bibitem [{\citenamefont {Kibler}\ \emph {et~al.}(2010)\citenamefont {Kibler},
  \citenamefont {Fatome}, \citenamefont {Finot}, \citenamefont {Millot},
  \citenamefont {Dias}, \citenamefont {Genty}, \citenamefont {Akhmediev},\ and\
  \citenamefont {Dudley}}]{Kibler2010Peregrine}%
  \BibitemOpen
  \bibfield  {author} {\bibinfo {author} {\bibfnamefont {B.}~\bibnamefont
  {Kibler}}, \bibinfo {author} {\bibfnamefont {J.}~\bibnamefont {Fatome}},
  \bibinfo {author} {\bibfnamefont {C.}~\bibnamefont {Finot}}, \bibinfo
  {author} {\bibfnamefont {G.}~\bibnamefont {Millot}}, \bibinfo {author}
  {\bibfnamefont {F.}~\bibnamefont {Dias}}, \bibinfo {author} {\bibfnamefont
  {G.}~\bibnamefont {Genty}}, \bibinfo {author} {\bibfnamefont {N.~N.}\
  \bibnamefont {Akhmediev}}, \ and\ \bibinfo {author} {\bibfnamefont {J.~M.}\
  \bibnamefont {Dudley}},\ }\href {\doibase 10.1038/nphys1740} {\bibfield
  {journal} {\bibinfo  {journal} {Nat. Phys}\ }\textbf {\bibinfo {volume}
  {6}},\ \bibinfo {pages} {790} (\bibinfo {year} {2010})}\BibitemShut {NoStop}%
\bibitem [{\citenamefont {Chabchoub}\ \emph {et~al.}(2011)\citenamefont
  {Chabchoub}, \citenamefont {Hoffmann},\ and\ \citenamefont
  {Akhmediev}}]{Chabchoub2011Rogue}%
  \BibitemOpen
  \bibfield  {author} {\bibinfo {author} {\bibfnamefont {A.}~\bibnamefont
  {Chabchoub}}, \bibinfo {author} {\bibfnamefont {N.~P.}\ \bibnamefont
  {Hoffmann}}, \ and\ \bibinfo {author} {\bibfnamefont {N.~N.}\ \bibnamefont
  {Akhmediev}},\ }\href {\doibase 10.1103/PhysRevLett.106.204502} {\bibfield
  {journal} {\bibinfo  {journal} {Phys. Rev. Lett.}\ }\textbf {\bibinfo
  {volume} {106}},\ \bibinfo {pages} {204502} (\bibinfo {year}
  {2011})}\BibitemShut {NoStop}%
\bibitem [{\citenamefont {Akhmediev}\ \emph {et~al.}(2009)\citenamefont
  {Akhmediev}, \citenamefont {Ankiewicz},\ and\ \citenamefont
  {Taki}}]{Akhmediev2009Waves}%
  \BibitemOpen
  \bibfield  {author} {\bibinfo {author} {\bibfnamefont {N.~N.}\ \bibnamefont
  {Akhmediev}}, \bibinfo {author} {\bibfnamefont {A.}~\bibnamefont
  {Ankiewicz}}, \ and\ \bibinfo {author} {\bibfnamefont {M.}~\bibnamefont
  {Taki}},\ }\href {\doibase 10.1016/j.physleta.2008.12.036} {\bibfield
  {journal} {\bibinfo  {journal} {Phys. Lett. A}\ }\textbf {\bibinfo {volume}
  {373}},\ \bibinfo {pages} {675} (\bibinfo {year} {2009})}\BibitemShut
  {NoStop}%
\bibitem [{\citenamefont {Bertola}\ and\ \citenamefont
  {Tovbis}(2013)}]{Bertola2013Universality}%
  \BibitemOpen
  \bibfield  {author} {\bibinfo {author} {\bibfnamefont {M.}~\bibnamefont
  {Bertola}}\ and\ \bibinfo {author} {\bibfnamefont {A.}~\bibnamefont
  {Tovbis}},\ }\href {\doibase 10.1002/cpa.21445} {\bibfield  {journal}
  {\bibinfo  {journal} {Commun. Pure Appl. Math.}\ }\textbf {\bibinfo {volume}
  {66}},\ \bibinfo {pages} {678} (\bibinfo {year} {2013})}\BibitemShut
  {NoStop}%
\bibitem [{\citenamefont {Tikan}\ \emph {et~al.}(2017)\citenamefont {Tikan},
  \citenamefont {Billet}, \citenamefont {El}, \citenamefont {Tovbis},
  \citenamefont {Bertola}, \citenamefont {Sylvestre}, \citenamefont {Gustave},
  \citenamefont {Randoux}, \citenamefont {Genty}, \citenamefont {Suret},\ and\
  \citenamefont {Dudley}}]{Tikan2017Universality}%
  \BibitemOpen
  \bibfield  {author} {\bibinfo {author} {\bibfnamefont {A.}~\bibnamefont
  {Tikan}}, \bibinfo {author} {\bibfnamefont {C.}~\bibnamefont {Billet}},
  \bibinfo {author} {\bibfnamefont {G.}~\bibnamefont {El}}, \bibinfo {author}
  {\bibfnamefont {A.}~\bibnamefont {Tovbis}}, \bibinfo {author} {\bibfnamefont
  {M.}~\bibnamefont {Bertola}}, \bibinfo {author} {\bibfnamefont
  {T.}~\bibnamefont {Sylvestre}}, \bibinfo {author} {\bibfnamefont
  {F.}~\bibnamefont {Gustave}}, \bibinfo {author} {\bibfnamefont
  {S.}~\bibnamefont {Randoux}}, \bibinfo {author} {\bibfnamefont
  {G.}~\bibnamefont {Genty}}, \bibinfo {author} {\bibfnamefont
  {P.}~\bibnamefont {Suret}}, \ and\ \bibinfo {author} {\bibfnamefont {J.~M.}\
  \bibnamefont {Dudley}},\ }\href {\doibase 10.1103/PhysRevLett.119.033901}
  {\bibfield  {journal} {\bibinfo  {journal} {Phys. Rev. Lett.}\ }\textbf
  {\bibinfo {volume} {119}},\ \bibinfo {pages} {033901} (\bibinfo {year}
  {2017})}\BibitemShut {NoStop}%
\bibitem [{\citenamefont {Tikan}\ \emph {et~al.}(2021)\citenamefont {Tikan},
  \citenamefont {Randoux}, \citenamefont {El}, \citenamefont {Tovbis},
  \citenamefont {Copie},\ and\ \citenamefont {Suret}}]{Tikan2021Local}%
  \BibitemOpen
  \bibfield  {author} {\bibinfo {author} {\bibfnamefont {A.}~\bibnamefont
  {Tikan}}, \bibinfo {author} {\bibfnamefont {S.}~\bibnamefont {Randoux}},
  \bibinfo {author} {\bibfnamefont {G.}~\bibnamefont {El}}, \bibinfo {author}
  {\bibfnamefont {A.}~\bibnamefont {Tovbis}}, \bibinfo {author} {\bibfnamefont
  {F.}~\bibnamefont {Copie}}, \ and\ \bibinfo {author} {\bibfnamefont
  {P.}~\bibnamefont {Suret}},\ }\href {\doibase 10.3389/fphy.2020.599435}
  {\bibfield  {journal} {\bibinfo  {journal} {Frontiers in Physics}\ }\textbf
  {\bibinfo {volume} {8}},\ \bibinfo {pages} {561} (\bibinfo {year}
  {2021})}\BibitemShut {NoStop}%
\bibitem [{\citenamefont {Tikan}(2020)}]{Tikan2020Effect}%
  \BibitemOpen
  \bibfield  {author} {\bibinfo {author} {\bibfnamefont {A.}~\bibnamefont
  {Tikan}},\ }\href@noop {} {\bibfield  {journal} {\bibinfo  {journal} {Phys.
  Rev. E}\ }\textbf {\bibinfo {volume} {101}},\ \bibinfo {pages} {012209}
  (\bibinfo {year} {2020})}\BibitemShut {NoStop}%
\bibitem [{\citenamefont {Michel}\ \emph {et~al.}(2020)\citenamefont {Michel},
  \citenamefont {Bonnefoy}, \citenamefont {Ducrozet}, \citenamefont
  {Prabhudesai}, \citenamefont {Cazaubiel}, \citenamefont {Copie},
  \citenamefont {Tikan}, \citenamefont {Suret}, \citenamefont {Randoux},\ and\
  \citenamefont {Falcon}}]{Michel2020Emergence}%
  \BibitemOpen
  \bibfield  {author} {\bibinfo {author} {\bibfnamefont {G.}~\bibnamefont
  {Michel}}, \bibinfo {author} {\bibfnamefont {F.}~\bibnamefont {Bonnefoy}},
  \bibinfo {author} {\bibfnamefont {G.}~\bibnamefont {Ducrozet}}, \bibinfo
  {author} {\bibfnamefont {G.}~\bibnamefont {Prabhudesai}}, \bibinfo {author}
  {\bibfnamefont {A.}~\bibnamefont {Cazaubiel}}, \bibinfo {author}
  {\bibfnamefont {F.}~\bibnamefont {Copie}}, \bibinfo {author} {\bibfnamefont
  {A.}~\bibnamefont {Tikan}}, \bibinfo {author} {\bibfnamefont
  {P.}~\bibnamefont {Suret}}, \bibinfo {author} {\bibfnamefont
  {S.}~\bibnamefont {Randoux}}, \ and\ \bibinfo {author} {\bibfnamefont
  {E.}~\bibnamefont {Falcon}},\ }\href {\doibase
  10.1103/PhysRevFluids.5.082801} {\bibfield  {journal} {\bibinfo  {journal}
  {Phys. Rev. Fluids}\ }\textbf {\bibinfo {volume} {5}},\ \bibinfo {pages}
  {082801} (\bibinfo {year} {2020})}\BibitemShut {NoStop}%
\bibitem [{\citenamefont {Goullet}\ and\ \citenamefont
  {Choi}(2011)}]{goullet2011numerical}%
  \BibitemOpen
  \bibfield  {author} {\bibinfo {author} {\bibfnamefont {A.}~\bibnamefont
  {Goullet}}\ and\ \bibinfo {author} {\bibfnamefont {W.}~\bibnamefont {Choi}},\
  }\href@noop {} {\bibfield  {journal} {\bibinfo  {journal} {Physics of
  Fluids}\ }\textbf {\bibinfo {volume} {23}},\ \bibinfo {pages} {016601}
  (\bibinfo {year} {2011})}\BibitemShut {NoStop}%
\bibitem [{\citenamefont {Bonnefoy}\ \emph {et~al.}(2020)\citenamefont
  {Bonnefoy}, \citenamefont {Tikan}, \citenamefont {Copie}, \citenamefont
  {Suret}, \citenamefont {Ducrozet}, \citenamefont {Prabhudesai}, \citenamefont
  {Michel}, \citenamefont {Cazaubiel}, \citenamefont {Falcon}, \citenamefont
  {El},\ and\ \citenamefont {Randoux}}]{Bonnefoy2020modulational}%
  \BibitemOpen
  \bibfield  {author} {\bibinfo {author} {\bibfnamefont {F.}~\bibnamefont
  {Bonnefoy}}, \bibinfo {author} {\bibfnamefont {A.}~\bibnamefont {Tikan}},
  \bibinfo {author} {\bibfnamefont {F.}~\bibnamefont {Copie}}, \bibinfo
  {author} {\bibfnamefont {P.}~\bibnamefont {Suret}}, \bibinfo {author}
  {\bibfnamefont {G.}~\bibnamefont {Ducrozet}}, \bibinfo {author}
  {\bibfnamefont {G.}~\bibnamefont {Prabhudesai}}, \bibinfo {author}
  {\bibfnamefont {G.}~\bibnamefont {Michel}}, \bibinfo {author} {\bibfnamefont
  {A.}~\bibnamefont {Cazaubiel}}, \bibinfo {author} {\bibfnamefont
  {E.}~\bibnamefont {Falcon}}, \bibinfo {author} {\bibfnamefont
  {G.}~\bibnamefont {El}}, \ and\ \bibinfo {author} {\bibfnamefont
  {S.}~\bibnamefont {Randoux}},\ }\href {\doibase
  10.1103/PhysRevFluids.5.034802} {\bibfield  {journal} {\bibinfo  {journal}
  {Phys. Rev. Fluids}\ }\textbf {\bibinfo {volume} {5}},\ \bibinfo {pages}
  {034802} (\bibinfo {year} {2020})}\BibitemShut {NoStop}%
\bibitem [{\citenamefont {Ryczkowski}\ \emph {et~al.}(2018)\citenamefont
  {Ryczkowski}, \citenamefont {N{\"a}rhi}, \citenamefont {Billet},
  \citenamefont {Merolla}, \citenamefont {Genty},\ and\ \citenamefont
  {Dudley}}]{ryczkowski2018real}%
  \BibitemOpen
  \bibfield  {author} {\bibinfo {author} {\bibfnamefont {P.}~\bibnamefont
  {Ryczkowski}}, \bibinfo {author} {\bibfnamefont {M.}~\bibnamefont
  {N{\"a}rhi}}, \bibinfo {author} {\bibfnamefont {C.}~\bibnamefont {Billet}},
  \bibinfo {author} {\bibfnamefont {J.-M.}\ \bibnamefont {Merolla}}, \bibinfo
  {author} {\bibfnamefont {G.}~\bibnamefont {Genty}}, \ and\ \bibinfo {author}
  {\bibfnamefont {J.~M.}\ \bibnamefont {Dudley}},\ }\href@noop {} {\bibfield
  {journal} {\bibinfo  {journal} {Nature Photonics}\ }\textbf {\bibinfo
  {volume} {12}},\ \bibinfo {pages} {221} (\bibinfo {year} {2018})}\BibitemShut
  {NoStop}%
\bibitem [{\citenamefont {Chekhovskoy}\ \emph {et~al.}(2019)\citenamefont
  {Chekhovskoy}, \citenamefont {Shtyrina}, \citenamefont {Fedoruk},
  \citenamefont {Medvedev},\ and\ \citenamefont
  {Turitsyn}}]{Chekhovskoy2019Nonlinear}%
  \BibitemOpen
  \bibfield  {author} {\bibinfo {author} {\bibfnamefont {I.~S.}\ \bibnamefont
  {Chekhovskoy}}, \bibinfo {author} {\bibfnamefont {O.~V.}\ \bibnamefont
  {Shtyrina}}, \bibinfo {author} {\bibfnamefont {M.~P.}\ \bibnamefont
  {Fedoruk}}, \bibinfo {author} {\bibfnamefont {S.~B.}\ \bibnamefont
  {Medvedev}}, \ and\ \bibinfo {author} {\bibfnamefont {S.~K.}\ \bibnamefont
  {Turitsyn}},\ }\href {\doibase 10.1103/PhysRevLett.122.153901} {\bibfield
  {journal} {\bibinfo  {journal} {Phys. Rev. Lett.}\ }\textbf {\bibinfo
  {volume} {122}},\ \bibinfo {pages} {153901} (\bibinfo {year}
  {2019})}\BibitemShut {NoStop}%
\bibitem [{\citenamefont {Sugavanam}\ \emph {et~al.}(2019)\citenamefont
  {Sugavanam}, \citenamefont {Kopae}, \citenamefont {Peng}, \citenamefont
  {Prilepsky},\ and\ \citenamefont {Turitsyn}}]{sugavanam2019analysis}%
  \BibitemOpen
  \bibfield  {author} {\bibinfo {author} {\bibfnamefont {S.}~\bibnamefont
  {Sugavanam}}, \bibinfo {author} {\bibfnamefont {M.~K.}\ \bibnamefont
  {Kopae}}, \bibinfo {author} {\bibfnamefont {J.}~\bibnamefont {Peng}},
  \bibinfo {author} {\bibfnamefont {J.~E.}\ \bibnamefont {Prilepsky}}, \ and\
  \bibinfo {author} {\bibfnamefont {S.~K.}\ \bibnamefont {Turitsyn}},\
  }\href@noop {} {\bibfield  {journal} {\bibinfo  {journal} {Nature
  communications}\ }\textbf {\bibinfo {volume} {10}},\ \bibinfo {pages} {1}
  (\bibinfo {year} {2019})}\BibitemShut {NoStop}%
\bibitem [{\citenamefont {Turitsyn}\ \emph {et~al.}(2020)\citenamefont
  {Turitsyn}, \citenamefont {Chekhovskoy},\ and\ \citenamefont
  {Fedoruk}}]{turitsyn2020nonlinear}%
  \BibitemOpen
  \bibfield  {author} {\bibinfo {author} {\bibfnamefont {S.~K.}\ \bibnamefont
  {Turitsyn}}, \bibinfo {author} {\bibfnamefont {I.~S.}\ \bibnamefont
  {Chekhovskoy}}, \ and\ \bibinfo {author} {\bibfnamefont {M.~P.}\ \bibnamefont
  {Fedoruk}},\ }\href@noop {} {\bibfield  {journal} {\bibinfo  {journal}
  {Optics Letters}\ }\textbf {\bibinfo {volume} {45}},\ \bibinfo {pages} {3059}
  (\bibinfo {year} {2020})}\BibitemShut {NoStop}%
\bibitem [{\citenamefont {Madelung}(1927)}]{Madelung1927Quantum}%
  \BibitemOpen
  \bibfield  {author} {\bibinfo {author} {\bibfnamefont {E.}~\bibnamefont
  {Madelung}},\ }\href@noop {} {\bibfield  {journal} {\bibinfo  {journal}
  {Zeitschrift f{\"{u}}r Physik}\ }\textbf {\bibinfo {volume} {40}},\ \bibinfo
  {pages} {1} (\bibinfo {year} {1927})}\BibitemShut {NoStop}%
\bibitem [{\citenamefont {El}\ and\ \citenamefont
  {Hoefer}(2016)}]{El2016Dispersive}%
  \BibitemOpen
  \bibfield  {author} {\bibinfo {author} {\bibfnamefont {G.}~\bibnamefont
  {El}}\ and\ \bibinfo {author} {\bibfnamefont {M.~A.}\ \bibnamefont
  {Hoefer}},\ }\href {\doibase 10.1016/j.physd.2016.04.006} {\bibfield
  {journal} {\bibinfo  {journal} {Physica D}\ }\textbf {\bibinfo {volume}
  {333}},\ \bibinfo {pages} {11} (\bibinfo {year} {2016})}\BibitemShut
  {NoStop}%
\bibitem [{\citenamefont {Dubrovin}\ \emph {et~al.}(2009)\citenamefont
  {Dubrovin}, \citenamefont {Grava},\ and\ \citenamefont
  {Klein}}]{Dubrovin2009universality}%
  \BibitemOpen
  \bibfield  {author} {\bibinfo {author} {\bibfnamefont {B.}~\bibnamefont
  {Dubrovin}}, \bibinfo {author} {\bibfnamefont {T.}~\bibnamefont {Grava}}, \
  and\ \bibinfo {author} {\bibfnamefont {C.}~\bibnamefont {Klein}},\ }\href
  {\doibase 10.1007/s00332-008-9025-y} {\bibfield  {journal} {\bibinfo
  {journal} {J. Nonlinear Sci.}\ }\textbf {\bibinfo {volume} {19}},\ \bibinfo
  {pages} {57} (\bibinfo {year} {2009})}\BibitemShut {NoStop}%
\bibitem [{\citenamefont {Tovbis}\ and\ \citenamefont
  {Venakides}(2000)}]{tovbis_eigenvalue_2000}%
  \BibitemOpen
  \bibfield  {author} {\bibinfo {author} {\bibfnamefont {A.}~\bibnamefont
  {Tovbis}}\ and\ \bibinfo {author} {\bibfnamefont {S.}~\bibnamefont
  {Venakides}},\ }\href {\doibase 10.1016/S0167-2789(00)00126-3} {\bibfield
  {journal} {\bibinfo  {journal} {Physica D: Nonlinear Phenomena}\ }\textbf
  {\bibinfo {volume} {146}},\ \bibinfo {pages} {150} (\bibinfo {year}
  {2000})}\BibitemShut {NoStop}%
\bibitem [{\citenamefont {Yang}(2010)}]{Yang:2010}%
  \BibitemOpen
  \bibfield  {author} {\bibinfo {author} {\bibfnamefont {J.}~\bibnamefont
  {Yang}},\ }\href@noop {} {\emph {\bibinfo {title} {Nonlinear waves in
  integrable and nonintegrable systems}}},\ Vol.~\bibinfo {volume} {16}\
  (\bibinfo  {publisher} {Siam},\ \bibinfo {year} {2010})\BibitemShut {NoStop}%
\bibitem [{\citenamefont {Tikan}\ \emph {et~al.}(2018)\citenamefont {Tikan},
  \citenamefont {Bielawski}, \citenamefont {Szwaj}, \citenamefont {Randoux},\
  and\ \citenamefont {Suret}}]{Tikan2018Single}%
  \BibitemOpen
  \bibfield  {author} {\bibinfo {author} {\bibfnamefont {A.}~\bibnamefont
  {Tikan}}, \bibinfo {author} {\bibfnamefont {S.}~\bibnamefont {Bielawski}},
  \bibinfo {author} {\bibfnamefont {C.}~\bibnamefont {Szwaj}}, \bibinfo
  {author} {\bibfnamefont {S.}~\bibnamefont {Randoux}}, \ and\ \bibinfo
  {author} {\bibfnamefont {P.}~\bibnamefont {Suret}},\ }\href {\doibase
  10.1038/s41566-018-0113-8} {\bibfield  {journal} {\bibinfo  {journal} {Nat.
  Photonics}\ }\textbf {\bibinfo {volume} {12}},\ \bibinfo {pages} {228}
  (\bibinfo {year} {2018})}\BibitemShut {NoStop}%
\bibitem [{\citenamefont {Clamond}\ \emph {et~al.}(2006)\citenamefont
  {Clamond}, \citenamefont {Francius}, \citenamefont {Grue},\ and\
  \citenamefont {Kharif}}]{clamond2006long}%
  \BibitemOpen
  \bibfield  {author} {\bibinfo {author} {\bibfnamefont {D.}~\bibnamefont
  {Clamond}}, \bibinfo {author} {\bibfnamefont {M.}~\bibnamefont {Francius}},
  \bibinfo {author} {\bibfnamefont {J.}~\bibnamefont {Grue}}, \ and\ \bibinfo
  {author} {\bibfnamefont {C.}~\bibnamefont {Kharif}},\ }\href@noop {}
  {\bibfield  {journal} {\bibinfo  {journal} {European Journal of
  Mechanics-B/Fluids}\ }\textbf {\bibinfo {volume} {25}},\ \bibinfo {pages}
  {536} (\bibinfo {year} {2006})}\BibitemShut {NoStop}%
\bibitem [{\citenamefont {Dudley}\ \emph {et~al.}(2006)\citenamefont {Dudley},
  \citenamefont {Genty},\ and\ \citenamefont
  {Coen}}]{Dudley2006Supercontinuum}%
  \BibitemOpen
  \bibfield  {author} {\bibinfo {author} {\bibfnamefont {J.~M.}\ \bibnamefont
  {Dudley}}, \bibinfo {author} {\bibfnamefont {G.}~\bibnamefont {Genty}}, \
  and\ \bibinfo {author} {\bibfnamefont {S.}~\bibnamefont {Coen}},\ }\href
  {\doibase 10.1103/RevModPhys.78.1135} {\bibfield  {journal} {\bibinfo
  {journal} {Rev. Mod. Phys.}\ }\textbf {\bibinfo {volume} {78}},\ \bibinfo
  {pages} {1135} (\bibinfo {year} {2006})}\BibitemShut {NoStop}%
\bibitem [{\citenamefont {Kivshar}\ and\ \citenamefont
  {Malomed}(1989)}]{Kivshar1989Dynamics}%
  \BibitemOpen
  \bibfield  {author} {\bibinfo {author} {\bibfnamefont {Y.~S.}\ \bibnamefont
  {Kivshar}}\ and\ \bibinfo {author} {\bibfnamefont {B.~A.}\ \bibnamefont
  {Malomed}},\ }\href {\doibase 10.1103/RevModPhys.61.763} {\bibfield
  {journal} {\bibinfo  {journal} {Rev. Mod. Phys.}\ }\textbf {\bibinfo {volume}
  {61}},\ \bibinfo {pages} {763} (\bibinfo {year} {1989})}\BibitemShut
  {NoStop}%
\bibitem [{\citenamefont {Suret}\ \emph {et~al.}(2020)\citenamefont {Suret},
  \citenamefont {Tikan}, \citenamefont {Bonnefoy}, \citenamefont {Copie},
  \citenamefont {Ducrozet}, \citenamefont {Gelash}, \citenamefont
  {Prabhudesai}, \citenamefont {Michel}, \citenamefont {Cazaubiel},
  \citenamefont {Falcon}, \citenamefont {El},\ and\ \citenamefont
  {Randoux}}]{Suret2020Nonlinear}%
  \BibitemOpen
  \bibfield  {author} {\bibinfo {author} {\bibfnamefont {P.}~\bibnamefont
  {Suret}}, \bibinfo {author} {\bibfnamefont {A.}~\bibnamefont {Tikan}},
  \bibinfo {author} {\bibfnamefont {F.}~\bibnamefont {Bonnefoy}}, \bibinfo
  {author} {\bibfnamefont {F.}~\bibnamefont {Copie}}, \bibinfo {author}
  {\bibfnamefont {G.}~\bibnamefont {Ducrozet}}, \bibinfo {author}
  {\bibfnamefont {A.}~\bibnamefont {Gelash}}, \bibinfo {author} {\bibfnamefont
  {G.}~\bibnamefont {Prabhudesai}}, \bibinfo {author} {\bibfnamefont
  {G.}~\bibnamefont {Michel}}, \bibinfo {author} {\bibfnamefont
  {A.}~\bibnamefont {Cazaubiel}}, \bibinfo {author} {\bibfnamefont
  {E.}~\bibnamefont {Falcon}}, \bibinfo {author} {\bibfnamefont
  {G.}~\bibnamefont {El}}, \ and\ \bibinfo {author} {\bibfnamefont
  {S.}~\bibnamefont {Randoux}},\ }\href {\doibase
  10.1103/PhysRevLett.125.264101} {\bibfield  {journal} {\bibinfo  {journal}
  {Phys. Rev. Lett.}\ }\textbf {\bibinfo {volume} {125}},\ \bibinfo {pages}
  {264101} (\bibinfo {year} {2020})}\BibitemShut {NoStop}%
\bibitem [{\citenamefont {Mei}(1989)}]{mei1989applied}%
  \BibitemOpen
  \bibfield  {author} {\bibinfo {author} {\bibfnamefont {C.~C.}\ \bibnamefont
  {Mei}},\ }\href@noop {} {\emph {\bibinfo {title} {The applied dynamics of
  ocean surface waves}}},\ Vol.~\bibinfo {volume} {1}\ (\bibinfo  {publisher}
  {World scientific},\ \bibinfo {year} {1989})\BibitemShut {NoStop}%
\end{thebibliography}%

\end{document}